\input epsf.sty
\documentclass{aa}
\begin{document}

\def\cs{c_{\rm s}}
\def\Rtr{R_{\rm tr}}
\def\rtr{r_{\rm tr}}
\def\Ti{T_{\rm i}}
\def\Te{T_{\rm e}}
\def\Ts{T_{\rm s}}
\def\Tv{T_{\rm v}}
\def\Fadv{F_{\rm adv}}
\def\Fsoft{F_{\rm soft}}
\def\Fcond{F_{\rm cond}}
\def\mdotz{\dot m_{\rm z}}
\def\mH{m_{\rm H}}
\def\me{m_{\rm e}}
\def\OmegaK{\Omega_{\rm K}}
\def\Prad{P_{\rm rad}}
\def\Ptot{P_{\rm tot}}
\def\Pgas{P_{\rm gas}}
\def\Pbase{P_{\rm base}}
\def\Revac{R_{\rm evac}}
\def\Rec10{R_{\rm evac}}
\def\Revap{R_{\rm evap}}
\def\tauevap{\tau_{\rm evap}}
\def\tauv{\tau_{\rm visc}}
\def\kappaes{\kappa_{\rm es}}
\def\kappao{\kappa_{\rm o}}
\def\alphac{\alpha}
\def\Qi{Q^{+}_{\rm i}}
\def\Q+e{Q^{+}_{\rm e}}
\def\Ftot{F_{\rm tot}}
\def\Fdisk{F_{\rm disk}}
\def\fdisk{f_{\rm disk}}
\def\fcor{f_{\rm cor}}
\def\Fcor{F_{\rm cor}}
\def\Fsoft{F_{\rm soft}}
\def\Fd{F_{\rm d}}
\def\RS{R_{\rm Schw}}
\def\rmax{r_{\rm max}}
\def\nuei{\nu_{\rm ei}}
\def\Fcool{F_{\rm cool}}
\def\taues{\tau_{\rm es}}
\def\MSun{{\rm M}_{\odot}}
\def\Hc{H_{\rm c}}
\def\const{{\rm const}}
\def\lambdaF{\lambda_{\rm F}}
\def\mcor{\dot m_{\rm cor}}
\def\mc26{\dot M_{\rm cor26}}
\def\mdotevapbranch{\dot m_{\rm evap.branch}}

   \thesaurus{03      
              (02.18.7;  
	       02.01.2;  
               11.01.2;  
	       11.17.3;  
	       11.19.1;  
	       13.25.2;  
               13.25.5)}  

   \title{Vertical structure of the accreting two-temperature corona and the transition to an ADAF}
\titlerunning{Disk/corona flow and transition to ADAF}


   \author{A. R\' o\. za\' nska
          \inst{1}
           \and
          B. Czerny\inst{1}
          }

   \offprints{A. R\' o\. za\' nska}

   \institute{Copernicus Astronomical Center, Bartycka 18, 00-716 Warsaw, Poland\\
             email: agata@camk.edu.pl
             }

   \date{Received ...; accepted ...}

   \maketitle

   \begin{abstract}

We investigate the model of the disc/corona accretion flow 
around  the black hole. 
Hot accreting advective corona is described by the two-temperature plasma in 
pressure
equilibrium with the cold disk. Corona is powered by accretion but it also 
exchanges energy with the disk through the radiative interaction and 
conduction. The model, parameterized by the
total (i.e. disk plus corona) accretion rate, $\dot m$ 
and the viscosity 
parameter, $\alpha$, uniquely determines the 
fraction of energy released in the corona as a function of radius and, in 
particular, the transition radius to the single-phase flow. 


Self-consistent solutions with the mass exchange
between phases display radial dependence of the parameters qualitatively 
different from the 'static' case, without the mass exchange. Corona covers 
the entire disk. The character of the radial dependence of the fraction of 
energy dissipated in the corona is qualitatively different for low and 
high total accretion rate. 
 
If the total accretion rate is low,   
the corona becomes stronger towards the central object, and finally the disc 
completely
evaporates, changing the accretion pattern into the single hot 
advection-dominated accretion flow (ADAF).
For intermediate accretion rates the reverse process - condensation - becomes
important, allowing 
possibly for a 
secondary disc rebuilding in the innermost part of the system.
High accretion rates always prevent the transition into ADAF, and the cold 
disk extends down to the marginally stable orbit.

The transition radius, $r_{tr}$, between the outer, two-phase flow and 
the inner, single-phase, optically thin flow, is equal to $4.51 \dot m^{-4/3}
 \alpha_{0.1}^7 R_{Schw}$ for $\dot m <6.9 \times 10^{-2} \alpha_{0.1}^{3.3}$
 and then contracts to the marginally stable orbit in a discontinuous way 
above this critical value of $\dot m$..  

This model reproduces all characteristic luminosity states of accretion 
black hole without any additional ad hoc assumptions. In particular.
the mechanism of the disk evaporation leads to a new, almost
horizontal branch on the accretion flow's stability curve (i.e. the dependence
of accretion rate on surface density) at the critical accretion rate.
This branch, together with the upper, advection dominated branch for 
optically thick disks, form boundaries for the time evolution of unstable, 
radiation pressure dominated disk. Therefore the disk at high accretion rates,
corresponding to Very High State in GBH and perhaps to Narrow Line Seyfert 1,
and quasar stage may oscillate between the disk dominated state and the 
evaporation branch state, with only a weak contribution from the cold disk 
emission.
The position of this branch for $\alpha = 0.1$ with respect to the gas 
pressure dominated branch is consistent with the presence of only weakly 
variable High State in GBH and the absence of a similar state in AGN: 
all the quasars vary considerably if monitored in timescales of years.
We also suggest a new interpretation of the Intermediate State, 
consistent with the presence of the strong reflected component.

  
   \end{abstract}

%

\section{Introduction}

\begin{figure}
\epsfxsize=8.8cm \epsfbox{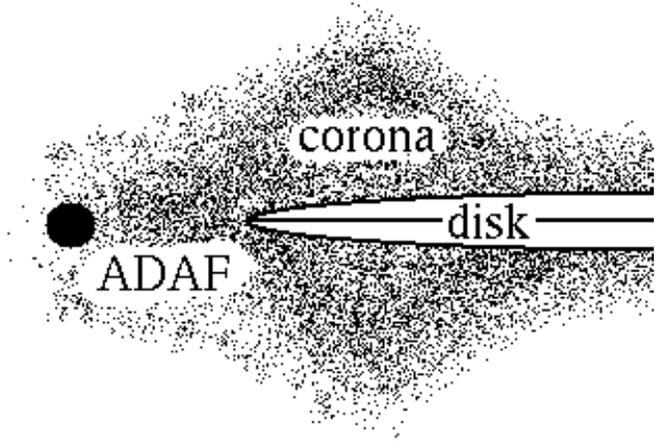}
\caption{The crossection through the central region of
accretion flow into a black hole.}
\label{fig:pict}
\end{figure}

There are strong observational evidence for the existence of the 
hot plasma $\sim 10^9$K in the neighborhood of the accretion disc
in active galactic nuclei (AGN) and galactic black holes (GBH).
Soft disc photons are Compton upscattered by this 
optically thin plasma and
form an approximately power law hard X-ray spectral component.
The presence of the Compton
reflection hump and the cold iron $K_{\alpha}$ line in the hard spectrum 
is a signature of strong illumination of cold disc matter by X-ray radiation 
(for a review see Mushotzky, Done \& Pounds 1993, Tanaka \& Lewin 1995, 
Madejski 1999).

The location of the hot plasma  is still under debate, but
one of the attractive models postulated by
Liang \& Price (1977) and Paczy\' nski (1978) 
introduces hot dissipative corona above 
accretion disc following the example of the solar corona.
via magnetohydrodynamic waves, jets or other nonthermal energy transport, 
and then thermalized in corona. But such ways of energy transport occurred 
to be inefficient, only $10 \%$ of energy generated in the disc can be 
carried out towards corona. 
Within the frame of such geometry, observations seem to require that in some 
sources and/or some spectral states the 
innermost part of the disk is either disrupted and replaced by an optically 
thin flow or there is a strong coronal outflow (e.g. Esin, McClintock 
\& Narayan 1997, Zdziarski, Lubi\' nski \& Smith 1999, Beloborodov 1999). 
Therefore, it is important to determine whether any of such effects are
expected on the basis of the physical description of the accretion flow 
(Narayan 1997).

The disc/corona models were recently investigated 
following two  basic lines.

The first group of models
does not explain the mechanism which heats up the 
plasma  in the disk/corona part of the flow. 
The fraction of energy generated in the corona $\fcor$
is treated as a free parameter of the model
(Haardt \& Maraschi 1991, Svensson \& Zdziarski 1994)
or changes according to the artificial formula being inversely 
proportional to  the distance form a central object (Esin et al. 1997). 
By changing $\fcor$ and the geometrical shape of the hot region
(continuous, patchy or clumpy corona),
such models can be fitted to the observed X-ray spectra 
(for review see Pountanen 1999). 
The main conclusion from those models is that static 
corona agrees with observations only when it does not cover the whole disc. 
However, more accurate considerations of the reflection, taking into 
account the optical depth of the Compton heated zone 
(R\' o\. za\' nska \& Czerny 1996, 
Nayakshin, Kazanas \& Kallman  1999) or an outflow
(Beloborodov 1999) may influence this result to some extent. 

The second group of models follow Shapiro, Lightman \& Eardley (1976) 
solution when hot optically thin flow exists in the inner part of the 
cold disc. They assume both cold and hot flow to be driven via 
accretion and depending on the model 
they take into account mechanical energy transport in the radial direction:
like advection (Ichimaru 1977;  Narayan \& Yi 1994, 1995a, 1995b;
Abramowicz et al. 1995; Chen et al. 1995;  for
review see Narayan, Mahadevan \& Quataert 1998  and Kato,
Fukue \& Mineshige 1998), turbulent diffusion
(Honma 1996), hydrodynamical inflow/outflow (Narayan \& Yi 1994, 1995a;
Blandford \& Begelman 1999)
and/or in the vertical direction: thermal conduction, 
evaporation  or coronal wind
(Meyer \&  Meyer-Hofmeister 1994 hereafter MMH94,
Meyer-Hofmeister \& Meyer 1999 Liu et. al 1999, 
Turolla \& Dullemond 2000). 

In both groups of models the crucial thing is the transition between the 
hot and the cold phase. In general picture such a transition  occurs in 
two directions: vertical transition when corona accretes above a cold disc,
and radial transition when disk/corona system is replaced with a 
single optically thin flow close to the black hole (Fig.~\ref{fig:pict}). 
Therefore, as was pointed by MMH94, in general picture 
both radial and vertical structure of the hot and cold phase should 
be considered. 

In this paper we present the simplest disc/corona model which is 
the combination of two groups of models described above.
The similar model was presented by Esin et al. (1997) with 
better description of ADAF region, and with non-local radiative interaction
between the hot  flow and cold  disc. Nevertheless, they
were not able to determine self consistently the transition between
disc/corona flow, 
since they did not calculate its vertical structure. 
Also MMH94 stressed that thermal conduction and
mass exchange between disc and corona in vertical direction are important
do determine the  transition. However, in their model
(Meyer-Hofmeister \& Meyer 1999 Liu et. al 1999) the 
radiative interaction between phases was neglected, an assumption of a 
single-temperature plasma was used, and 
the conductive flux on the bottom of the hot phase was
artificially taken from the solar corona.

Our model takes into account
radiative coupling between phases, thermal conduction and evaporation 
in vertical direction, and radial energy transport via advection.
The mass exchange rate is calculated self-consistently from the
physical conditions in the transition zone.
Hot corona is treated as two temperature plasma and, similarly to the disc,
it is powered by accretion 
(Witt, Czerny \& \.Zycki 1997; hereafter WC\. Z97, 
Esin et al. 1997).

On each radius the vertical structure is calculated 
depending on comparison  of accretion and vertical evaporation timescales.
When accretion time scale is long enough, the mass exchange can be
neglected and the 'static' equilibrium is achieved.
In such a case, for vertical transition we use the condition based on 
considerations of Krolik (1998), Dullemond (1999), and
R\' o\. za\' nska \& Czerny (2000) which is relevant to the dissipative 
corona. When accretion time scale is short  mass exchange cannot be neglected,
and we determine the evaporation rate through the disc/corona transition.

Vertical properties are used to compute the radial trends solving continuity 
equation.   
The aim of this paper is to determine the radial dependence of the fraction
of energy liberated in the corona for a given accretion rate, viscosity 
and a mass
of the black hole and to check when the disk evaporation leads to a transition
from disc/corona flow to the single optically thin flow.
The radial transition occurs at the radius when all the mass starts  to 
be carried by the corona. 

Observational consequences of our model occur in the dependence of the 
transition radius on accretion rate, which can be seen in different shapes of
the reflected component (Done \& \. Zycki 1999).  
Also the evaporation leads to the new stable branch on
the accretion flow's stability curve ( i.e. $\dot m ( \Sigma)$), allowing for
new evolution cycle to exist. If accretion rate is too low optically
thick disk can evaporate completely in the innermost parts
and the position of transition radius moves outward. This can explain the
different luminosity
states in AGNs and GBHs   

In Sec \ref{sec:mod} we present the model. Calculation are shown in Sec. 
\ref{sec:res}, and we discuss our results in Sec. \ref{sec:dis}. 
Evaporation rate is determined
in Appendix A.

\section{Model}
\label{sec:mod}
The basic problem in modeling the two-phase accretion flow onto a black 
hole is the determination of the relative importance of the hot optically
thin flow and cold optically thick flow. It was shown by 
MMH94 in the case of dwarf novae that such
estimates have to be based on the study of the vertical structure of such
a flow since only in that case we can determine the mass exchange between
the phases and the conditions of their coexistence. Therefore we follow this
basic approach but we adjust the physical input to the conditions appropriate
for the accretion close to the black hole. 

The main difference from a dwarf novae is that in the case of  
inner regions of the accretion disc in AGNs and GBH, two-temperature plasma
should be taken into account (Narayan \& Yi 1995b).
This is easy to see from the following simple argument.
With the assumption of local viscous heating of ions, the total flux    
generated  locally in  the standard disk is:
\begin{equation}
F_{\rm gen}= {3 \over 2} \alpha P \OmegaK H,
\label{eq:gen}
\end{equation} 
where $\OmegaK$ is the Keplerian angular frequency equal to $(GM/r^3)^{1/2}$,
$\alpha$ is the viscosity, $H$ is the height scale of the flow and P is 
the pressure.

On the other hand, the total flux dissipated locally in a stationary 
Keplerian disk can be described as: 
\begin{equation} 
F_{\rm gen}= {3 G M \dot M \over 8 \pi r^3} (1 - f(r/\RS)),
\label{eq:diss}
\end{equation}
where $f(r/\RS)$ represent the inner boundary condition, in our case taken
from the Newtonian approximation, $f(r/\RS)= (3\RS/r)^{1/2}$. 

The optical depth of the flow can be described by 
$\tau_{es}=\kappa_{es} \rho H$,
where $\kappa_{es}$ is the opacity coefficient for electron scattering 
and $\rho$ is the density. 
We use  also following  equation of state: $P={k \over  \mu \mH} \rho T$, 
with the value of molecular weight $\mu = 0.5$ for cosmic
chemical composition. Boltzmann constant is denoted as $k$ and the mass of
hydrogen atom as $\mH$.

Combining these equations, we can estimate the temperature 
of the optically thin $\tau_{\rm es} \sim 1$ flow as:
\begin {equation}
T=2.31 \times 10^{13} {\dot m  \over \alpha}  (\RS/r)^{3/2}[1-(3\RS/r)^{1/2}] 
\end{equation}
We use accretion rate $\dot m = \dot M/\dot M_{Edd}$ in units of 
Eddington accretion rate described as:
\begin{equation} 
\dot M_{Edd} = \frac {4 \pi G M}{\eta \kappa_{es} c}
\label{eq:Edd}
\end{equation} 
with accretion efficiency $\eta =1/12 $.
Thus the temperature does not depend on the mass of the black hole $M$, 
and can be computed for assumed value of Eddington accretion rate, 
radius and viscosity.
Taking a realistic value $\dot m  / \alpha \sim 1$, and
the distance appropriate for dwarf nowae which is $10^4 \RS$ we obtain the 
temperature equal to $10^7$ K, reasonable for the single 
temperature plasma. But, for $r=10 \RS$, the temperature attains
the value of $4 \times 10^{11}$ K, far too high for electrons, according to 
the observational data, and two-temperature plasma should be considered.   

\subsection{Basic assumptions and model parameters}

In the present paper we consider the case of stationary accretion onto a 
non-rotating black hole. We assume that accreting material forms an
optically thick disk underlying an optically thin hotter skin -
a corona. The flow is described as a  continuous medium
so a possible clumpiness of either the disk or the corona are 
neglected. The transition of the flow in the vertical direction from
a corona to a disk is gradual and corresponds to a change of the cooling
mechanism with an increasing density and pressure towards the equatorial
plane. Since the computations of the optically thick part of the flow
required another approach than the optically thin parts we still introduce a
formal division of the flow into a disk and the corona but the
physical parameters at the bottom of the corona (pressure, density and
temperature) are equal to the disk values so there is no discontinuity
between the two media. 

We assume that both the disk and the corona are powered directly
by the release of the gravitational energy due to accretion. Viscous torque
is parameterized as in the standard disk of Shakura \& Sunyaev (1973). This
approach is supported to some extent by the magnetohydrodynamical simulations
of Hawley, Gammi \& Balbus (1996). Such assumption was used e.g. by MMH94 and
WC\. Z97. Since the corona is accreting we assume the 
direct heating of ions and the subsequent energy transfer to electrons through
the Coulomb coupling. Therefore, the corona is basically a two-temperature
medium unless the Coulomb coupling is efficient enough to equalize the 
temperatures. This assumption is a simple, well determined 
starting point but in the future
the ion-electron energy exchange should be considered more closely. 
Proton heated plasma, adopted in our paper, is hardly 
thermalised (Gruzinov \& Quataert 1999). Spruit \& Haardt (2000), motivated
by large mean free path of ions considered the disk heating not as conduction
due to electrons, but conduction due to the ions.
Another point is that protons are heated efficiently only in the case of a 
weakly magnetized plasma; in a
plasma with field in equipartition heat goes to electrons (Quataert \&
Gruzinov 1999). Also
Ohmic heating stressed by Bisnovatyi-Kogan \& Lovelace (1997)
can efficiently heat up the electrons and advection may be not important.
Diminishing the role of advection may have further consequences for the 
efficiency of mass exchange between the disk and the corona 
(R\' o\. za\' nska \& Czerny 2000). 

We take into account the radial advection, Compton and bremsstrahlung cooling
and the conduction due to electrons and the vertical mass exchange,
which are important at the basis of the 
corona. 
Disk and corona are radiatively coupled, as in the classical paper
of Haart \& Maraschi (1991) although the cooling process is now treated 
locally.  We assume that, the cold disc is the origin of the seed photons
at a given radius. We neglect the contribution of the synchrotron 
radiation to the seed photons for Comptonization (Narayan \& Yi 1994), 
as well as non-local photons which were included in the study of Esin et al.
(1997). We also neglect the formation of wind at the top of corona
discussed in some papers (e.g. Begelman, McKee \& Shields 1983, Begelman
\& Blandford 1999).

In flows cooled by advection, the motion may be not
Keplerian and proper calculation of equation of motion should be
done to describe this effect. Nevertheless, the aim of this paper
is to calculate properly the vertical structure while the radial one is 
computed less carefully. Therefore, together with  
hydrostatic equilibrium in vertical direction, we assume the 
Keplerian motion of the material. 

The model parameters are: mass of the black hole, $M$, accretion rate, 
$\dot m$, and the viscosity parameter $\alpha$, assumed here for simplicity 
to be the same both in the disk and in the corona. All other quantities,
including their radial dependences, are uniquely determined by those global
parameters for a stationary flow. Corona model does not
actually depend on $M$ but there is a weak dependence of the cold disk 
properties on the mass of the black hole itself so this parameter cannot be 
entirely neglected.

\subsection{Vertical structure of the corona at a given radius}

In the present study we simplify the problem of hydrostatic equilibrium. 
Instead
of solving complex differential equations as in MMH94, 
we simply study algebraic dependences of various quantities on the
pressure which is a monotonic function of the distance from equatorial plane.
With such assumption the local viscous heating of ions $\Qi$ 
can be described following the Eq. \ref{eq:gen}:
\begin{equation}
\Qi = {3 \over 2} \alpha P \OmegaK,
\label{eq:genloc}
\end{equation} 

A fraction of this energy is carried by advection. This fraction is 
proportional to the ratio of the ion temperature $\Ti$ to the virial 
temperature
$\Tv$ with the coefficient $\delta$ determined by the radial derivatives
of thermodynamical quantities 
(e.g. Ichimaru 1977, Muchotrzeb 
\& Paczy\' nski 1982, Abramowicz et al. 1995, Janiuk, \. Zycki \& 
Czerny 2000). At present
we simply assume $\delta=1$ but in principle this value should be determined
iteratively.

Remaining fraction of energy is transfered to electrons via Coulomb coupling
\begin{equation}
\Qi \left( 1 - \delta {\Ti \over \Tv} \right) = \Q+e = {3 \over 2} 
{k \over \mH} D \rho^2 (\Ti - \Te)\Te^{-3/2} 
\label{eq:coulomb}
\end{equation}
where $\Te$ is the electron temperature, and
D is the Coulomb coupling constant equal $4.88 \times 10^{22}$ (in s$^{-1}$).
At this exploratory stage, we neglect relativistic corrections to this
equation given by Stepney \& Guilbert (1983). 

In the upper layers of the corona the electrons cool down radiatively, i.e.
\begin{equation}
\Q+e= \Lambda (\rho,\Te). 
\label{eq:qool}
\end{equation}
The processes included are the Compton cooling due to the presence
of soft photons $\Fsoft$ from the disk and through the bremsstrahlung cooling
\begin{equation}
\Lambda (\rho,\Te) =  \Fsoft \kappaes { 4 k \over \me c^2} \rho \Te + B \rho^2 
\Te^{1/2}.
\label{eq:elcool} 
\end{equation}
Here $\Fsoft$ is the soft radiation flux coming from the disk due to the
energy generation in the optically thick part of the flow as well as due to
interception and reemission of the coronal emission 
$F_{\rm hard}$ (Haardt \& Maraschi 1991)
\begin{equation}
\Fsoft = \Fdisk + 0.5 {\bf F_{\rm hard}}, 
\end{equation} 
$\me$ - mass
of the electron, $c$ - light velocity, and $B$ - bremsstrahlung cooling 
constant equal $6.6 \times 10^{20}$ erg s$^{-1}$cm$^{-2}$ g$^{-2}$.
We assume that half of the coronal flux ${\bf F_{\rm hard}}$, 
illuminates the disc and all of
this is absorbed and subsequently reemited, so albedo is equal to zero. 

In Eq.~\ref{eq:elcool} we include only the Compton cooling and we neglect, for
simplicity, the Compton heating term which may be important within the transition zone where the electron temperature drops below $10^8$ K. 

Equations \ref{eq:coulomb} and \ref{eq:elcool} determine the temperature 
profile in the upper part of the corona, or more precisely, the 
dependence of the ion and electron temperature on the gas pressure which we
identify with the ion pressure for simplicity, $P=k/\mH \rho \Ti$.

At some depth, however, a transition to a cool disk is expected. Within such
a transition zone the essential role is played by conduction, neglected so
far. Here we assume that the transition from hot to cold gas happens very
sharply, under the constant pressure, $\Pbase$, so the drop in the 
temperature is
simply compensated by the increase in the density. Determination of this 
pressure is the key element of the model since the value of this pressure
uniquely determines the fraction of the energy $\fcor$ liberated in the 
corona in the following way.

Since the ion temperature is close to the virial temperature, the pressure
scale high is roughly of the order of the radius, $r$, so instead of solving
precisely the hydrostatic equilibrium we can estimate the total energy flux
generated in the corona as
\begin{equation}
\Fcor = \Qi (\Pbase) r,
\end{equation}
and the flux emitted by corona as
\begin{equation}
F_{\rm hard} = \Qi \left(1- \delta {\Ti \over \Tv} \right)  r.
\end{equation}
The total energy dissipated due to accretion both in the disk and the corona
is given by formula:
\begin{equation}
\Ftot=\Fcor + \Fdisk=\Fsoft + 0.5  F_{\rm hard} \frac{1+
\delta {\Ti \over \Tv}}{1-\delta {\Ti \over \Tv}},
\label{eq:sum}
\end{equation} 
which relates to the accretion rate both within the disk
and the corona due to the assumption of stationarity and Keplerian motion
and  $\Ftot=F_{\rm gen}$ given by Eq. \ref{eq:diss}.

In the accreting corona model, the fraction of energy dissipated in the 
corona corresponds to the same fraction of the mass accreting through the 
hot phase, i.e.
\begin{equation}
{\dot m_{\rm cor} \over  \dot m} = {\Fcor \over \Ftot}
\end{equation}

In this way the determination of the pressure at the basis of the corona
allows to determine its structure at a given radius from the global model
parameters, $M$, $\dot m$, and $\alpha$. Therefore the way how we determine
this value is a crucial element of the model itself which removes 
arbitrariness from the disk/corona accretion flow present in most models 
apart from MMH94, \. Zycki, Collin-Souffrin \& Czerny (1995),  WC\. Z97 
and the  directly related papers.

The computations of the vertical structure of the optically thick part of the
flow are not necessary for the computation of the corona. However, when the
corona structure is determined the disk structure is calculated, with
the boundary conditions imposed by the presence of corona, as described
by R\' o\. za\' nska et al. (1999). No iterations between the disk and the
corona computations are needed. 

The properties of the
transition zone are calculated following the approach of Krolik (1998), 
Dullemond (1999), and 
R\' o\. za\' nska
\& Czerny (2000). We consider two cases: a
'static' balance solution without the mass exchange between the disk and the
corona and the self-consistent case based on the continuity equation of
the coronal flow.  

\subsubsection{Equilibrium solutions without mass exchange}

Interesting and important solutions are obtained under an assumption that
the accretion flow is slow in comparison with the timescale of exchange
of the mass between the disk and the corona. In that case the equilibrium
solution is  defined by  the condition that there is no mass exchange between
the disk and the corona (see R\' o\. za\' nska \& Czerny 2000). 

To describe the equilibrium solution we integrate the energy balance 
equation as shown in Appendix A, with velocity in vertical direction
$v_{\rm z}$ equal zero  (see Eq. \ref{eq:eninteg}).
It meant, that the heat flux which is transported via electrons
from hot upper corona towards the cold matter is balanced by
radiative cooling operating in the disc.  
Since the transition is described under constant
pressure the appropriate criterion reduces to an integral 
(R\' o\. za\' nska \& Czerny 2000)
\begin{equation}
\int_{\Ts}^{\Te(P)} (\Q+e - \Lambda)\kappao \Te^{5/2} d\Te  = 0
\label{eq:noevap}
\end{equation}
where $\Ts$ is the cold disk temperature and $\Te (P)$ is the electron 
temperature within the upper corona part determined at the pressure P
and satisfying the heating/cooling balance formulated by 
Equation \ref{eq:qool}.
Here, $\kappa=\kappao T^{5/2}$ is thermal conductivity coefficient 
in ${\rm ergs \, cm}^{-1}{\rm s}^{-1} {\rm K}^{-1}$ and
conductivity constant $\kappao$ is assumed to be equal $5.6 \times 10^{-7}$.
The ion temperature in this equation 
has to be determined from the Coulomb coupling equation (Eq. \ref{eq:coulomb})
and the density from the value of the (constant) pressure and the ion
temperature.

Equation \ref{eq:noevap} determines the pressure of the transition zone, thus
closing up the set of equations determining the disk/corona flow.

\subsubsection{Solutions with evaporation/condensation}

The mass exchange between the disk and the corona cannot be neglected if the
timescale of accretion is short so the 'static' type of equilibrium cannot
develop. It is also necessary if we want to describe the process of corona
formation. This approach is more appropriate but more complex since it is
essentially non-local and requires a global solution for all radii. Here
we basically follow an approach of MMH94 by
finding a family of numerical solutions at a single radii, determining their
scaling properties with the radius and finally finding unique global solution 
at the basis of this scaling.

Therefore we allow the Equation \ref{eq:noevap} to be not satisfied which
means that we have to solve the vertical energy transfer in the transition
zone allowing for evaporation or condensation, depending on the choice of the
pressure $\Pbase$. 
If radiative processes are to low to balance the heat flux, the disc will
evaporate as shown by Dullemond (1999). However, radiative cooling may
overwhelm the heat flux if pressure at the base is high enough
(higher than the pressure under which  mass exchange equals to zero, 
as shown by  McKee \& Begelman in 1990), and in such situation matter 
will condense. 

To determine the mass exchange rate we integrate the
energy balance equation (\ref{eq:general}) in the way detaily presented in 
Appendix A. 
Combining equations (\ref{eq:lamfield}) and (\ref{eq:conpar}) the 
evaporation rate $\mdotz$ across the disk/corona boundary is given by the 
formula.
\begin{equation}
\mdotz = { \int_{\Ts}^{\Te(P)} [\Q+e - \Lambda]\kappao \Te^{5/2} d\Te  
\over  |\int_{\Ts}^{\Te(P)} (\Q+e - \Lambda)\kappao \Te^{5/2} d\Te|^{1/2}}
\left( {2 \over 5}{ \mH \over  k \Ti(P)} \right)
\end{equation}
The sign of this quantity is either positive (disk material evaporates into
the corona) or negative (coronal material condensates into the disk), 
depending on the net effect of the electron heating and cooling balance.

The radial dependence of $\Pbase$ is subsequently determined from the
continuity equation (see Section \ref{sect:evap}).

\subsection{Radial component of the continuity equation and 
global solutions}

\subsubsection{Solutions without mass exchange}

In this case the equilibrium is built locally so the determination of the 
disk/corona structure at a given radius is completed if the local solution at
this radius is found. The accretion rate within the corona is determined
at each radius separately and the mass exchange between the disk and the
corona account for its dependence on the disk radius, i.e. its evolution
with the flow. All parameters, like accretion rate in the corona, soft X-ray
flux from the disk, the ion and electron temperature profile, the density
profile and the corona optical depth as well as the pressure at the basis
of the corona where the temperature drops down to match the disk values are
determined by the choice of the radius $r$ and the three global model 
parameters, $M$, $\dot m$ and $\alpha$.

\subsubsection{Solutions with evaporation/condensation}
\label{sect:evap}

If the timescale of building the evaporation/condensation equilibrium is long
in comparison with the viscous timescale in the corona we have to describe
the build-up process of the corona, i.e. the accretion rate in the corona
at any radius $r$ will reflect the amount of material which evaporated from 
the disk at all radii larger and equal to $r$.  

The change of the accretion rate in the corona $\dot M_{\rm cor}$ is therefore due
to the mass exchange with the disk
\begin{equation}
{d\dot M_{\rm cor} \over dr} = - 2 \pi r \mdotz.
\label{eq:cont}
\end{equation}
which in dimensionless units reduces to
\begin{equation}
{d\mcor \over dR} = - \frac {2 \eta \kappa_{\rm es} G}{c^3}  M R \mdotz = - 27.9 
M_{\rm BH8} R \mdotz
\label{eq:cont}
\end{equation}
where $R=r/\RS$ and $M_{\rm BH8}=M/10^8 M_{\odot}$.

In order to solve this equation we follow the approach of MMH94.
 We find a family of solutions at a given radii for a set of
values of the pressure $\Pbase$. We express those solutions through the
computed $\mcor$, thus obtaining  a numerical relation $\mdotz (\mcor)$. We
study the scaling of this relation with radius, thus finding an analytical
expression valid in the entire disk. This analytical expression is supplied
to Equation~\ref{eq:cont} and the final analytical result for the dependence
$\mcor(r)$ is found.

\subsection{Cold disk structure}
\label{sect:cold}

Determination of the corona properties, i.e. the fraction of energy 
dissipated in the corona and the pressure at the corona basis provides the
boundary conditions necessary  to 
calculate the vertical structure of the cold disk flow 
(e.g. R\. o\. za\' nska et al. 1999). Therefore, the description of the hot 
flow influences the properties of the cold part of the flow to their strong
coupling through these boundary conditions.

In the present paper we concentrate mostly on the hot flow properties and the
cold disk properties are only required for the qualitative discussion of the 
connection 
between the single phase hot flow and disk/corona flow (see 
Section~\ref{sect:mdotsig}) 
so we simplify the description of the cold flow. We apply the vertically 
averaged approach but we determine the dimensionless coefficients involved in 
this process on the basis of full computations of the disk vertical structure,
as in Janiuk et al. (2000) (see also Muchotrzeb \& Paczy\' nski 1982,
Abramowicz et al. 1988). 

We assume the same value of the viscosity coefficient in the hot and in the 
cold flow although they may in principle be considerably different due to the
different plasma properties.  We include advection in our description of 
the flow, and the appropriate coefficients used are: $B_1=B_2=B_3=B_4=1$, 
and $q_{\rm adv}=2$.
The disk is assumed to be optically thick so the current approach does not 
allow yet to follow the final transition from the disk/corona flow to a 
single phase optically thin accretion.

\section{Results}
\label{sec:res}

\subsection{Solutions without mass exchange}

\label{sect:noevap}

\subsubsection{Vertical structure of the flow at a given radius}

We first discuss the vertical ion and electron temperature and density 
profiles at a representative radius $r=10.8 \RS $ for mass of the black hole
equal $10^8 \MSun$ and the viscosity parameter $\alpha = 0.1$. 
The virial temperature corresponding to this radius is
equal $10^{12}$ K.  

The set of algebraic equations (Eq.~\ref{eq:diss} and Eqs.~\ref{eq:genloc} - 
\ref{eq:sum}) describing the upper layers of the corona
formally posses two solutions for a given pressure (corresponding to a certain
distance $z$ from the equatorial plane): one solution is mostly advection-dominated, low
density branch and the other one is high density, mostly radiatively cooled
branch. This property, found also by Dullemond (1999), is a direct analogy
of the vertically averaged optically thin solutions found by Abramowicz et al.
(1995). Both solutions exists for low values of the pressure
but they approach each other and finally merge at a certain value of the
pressure, corresponding to an ion temperature $\Ti = 0.5 \Tv$. Dipper inside
there are no solutions, as in the case of too high accretion rate in optically
thin flows.

However, the situation of the disk/corona accretion is not the same as in the
case of a single phase flow. Only one of the above solutions is acceptable.
At the advection-dominated branch the ion temperature decreases with the 
increase towards the midplane and the density increase thus opening a possibility
to match this solution to the cool disk part of the flow. The other branch
shows opposite trends and therefore we reject it as unphysical in the case
of two-phase accretion.

The fact, that the algebraic solution does not extends very deeply does not
pose a problem since we have now to take into account the transition to the
disk flow under no evaporation conditions. The value of the pressure which 
allows the coexistence of the disk and coronal flow under no 
evaporation/condensation condition given by Equation~\ref{eq:noevap} is always
smaller than the value of the pressure at which the two solutions merge.

\begin{figure}
\epsfxsize=8.8cm \epsfbox{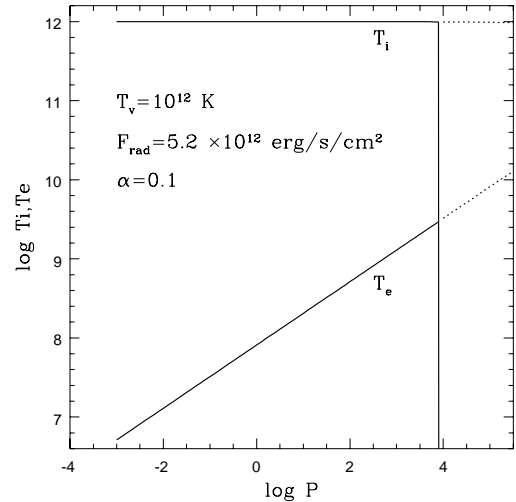}
\caption{The change of the ion and electron temperature with pressure
increasing towards equatorial plane for $\dot m=1.3 \times 10^{-3}$.
Vertical continuous line mark the 
$\Pbase$ value at which the temperature drops dramatically to match the disk
values without, however, mass exchange between the disk and the corona. 
In such a solution $\fcor=0.89$. If analogous  transition takes place at 
lower pressure it would lead to continuous disk evaporation due to
conduction while the delayed transition at higher pressure would result in
continuous corona condensation into a cold material and in each case 
$\fcor$ would take different value.}
\label{fig:Tpprof}
\end{figure}

The example of the change of the ion and electron temperature with a pressure
increasing towards the equatorial plane is shown in Fig.~\ref{fig:Tpprof}
for the accretion rate $\dot m =1.3 \times 10^{-3}$  in units of Eddington
accretion rate.
The ion temperature is very close to the virial value, 
most of the heat generated
in the corona is advected towards smaller radii. The electron temperature
is always significantly lower than the ion temperature since the Coulomb 
coupling is nor very efficient but it slowly increases towards the equatorial
plane (i.e. increasing $P$). For the assumed parameters the equilibrium pressure $\Pbase$ is equal
$7.94 \times 10^{3}$ g cm$^{-3}$. 

The fraction of energy
generated in the corona is equal 0.89, so the soft photon flux is predominantly
due to the fraction of corona emission absorbed by the disk ($\Fsoft= 5.2 \times
10^{12}$erg s$^{-1}$cm$^{-2}$). The optical depth of the corona is very low,
$6 \times 10^{-3}$ and the density at the upper edge of the transition 
zone is equal $5 \times 10^{-16}$ g s$^{-2}$ cm$^{-1}$.

\begin{figure}
\epsfxsize=8.8cm \epsfbox{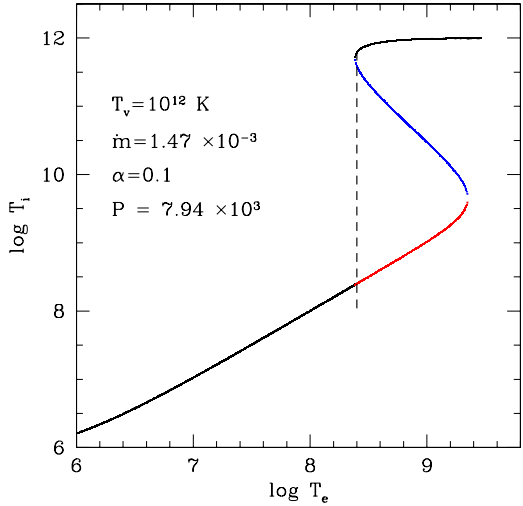}
\caption{The relation between the ion and electron temperature within 
transition zone.}
\label{fig:TTprof}
\end{figure}

The transition zone itself is also complex. We assumed that the transition 
happens at a constant pressure so its structure cannot be shown in 
Fig.~\ref{fig:Tpprof}. Instead, we draw the relation between the ion and
electron temperature within this zone, as predicted, neglecting thermal 
conduction (only for radiative processes and advection), in 
Fig.~\ref{fig:TTprof}.

In the upper part of the transition zone itself the ion temperature is still 
very close to the virial temperature, with advection as a dominant cooling 
mechanism 
for ions. However, deeper inside the ion temperature decreases and the density
increases. Radiative cooling becomes more efficient. The advection-dominated 
flow is replaced by the radiatively cooled flow, basically of the type studied
by Shapiro, Lightman \& Eardley (1976). This branch was found to be unstable
(Piran 1978, Narayan \& Yi 1995b) and it does not provide us with a monotonic 
decrease of the electron 
temperature so we assume that actually a sharp transition takes place along a
vertical line, as marked in Fig.~\ref{fig:TTprof}. Our integral of the
difference between the heating and cooling multiplied by the conduction flux
(Equation~\ref{eq:noevap}) is calculated applying such a profile. Full solution
of differential equations with the conduction due to electrons as well as
ions would produce a smooth solution matching the upper and lower branch
without the third, intermediate branch. We intend to study this problem in the
future but for the determination of the basis solution properties the detailed
knowledge of the transition zone is not necessary as the optical depth of this
zone is generally very small. 

\begin{figure}
\epsfxsize=8.8cm \epsfbox{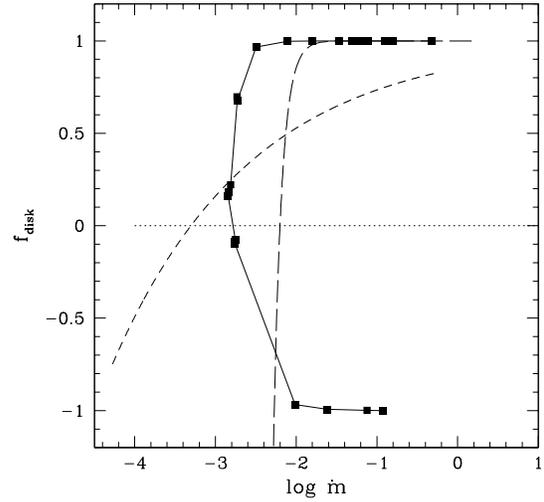}
\caption{The dependence of the fraction of energy generated in the disk 
$\fdisk=1 - \fcor$ on
the total accretion rate at $10 \RS$: equilibrium solution (continuous line
with squares), solution with evaporation/condensation (long dashed line)
and prediction from 
WC\. Z97 (short dashed line), Solutions below dotted line are 
unphysical, but they are shown to demostrate the mathematical reason
for the disappearance of solutions for low $\dot m$.}
\label{fig:fdisk10}
\end{figure}

The fraction of the energy generated in the corona for an accretion rate of
$\dot m = 1.47 \times 10^{-3}$  is high but generally this is not the case. 
In Fig.~\ref{fig:fdisk10}
we show the dependence of this value on the accretion rate at the radius
10 $\RS$. For large accretion rates the corona is weak and its strength 
increases rapidly when the accretion rates drops down to 
$ \dot m= 1.53 \times 10^{-3}$. Below $1.47 \times 10^{-3}$ 
there are no solutions for disk/corona equilibrium without the mass exchange.
Solutions with negative value of $f_{\rm disk}$ are
unphysical.
Those results are qualitatively very similar to the results obtained by 
WC\. Z97 which were based on approximate description of the disk/corona
equilibrium through the condition on the value of the ionization parameter 
$\Xi$ (shown in Fig.~\ref{fig:fdisk10} as a dashed line).

\begin{figure}
\epsfxsize=8.8cm \epsfbox{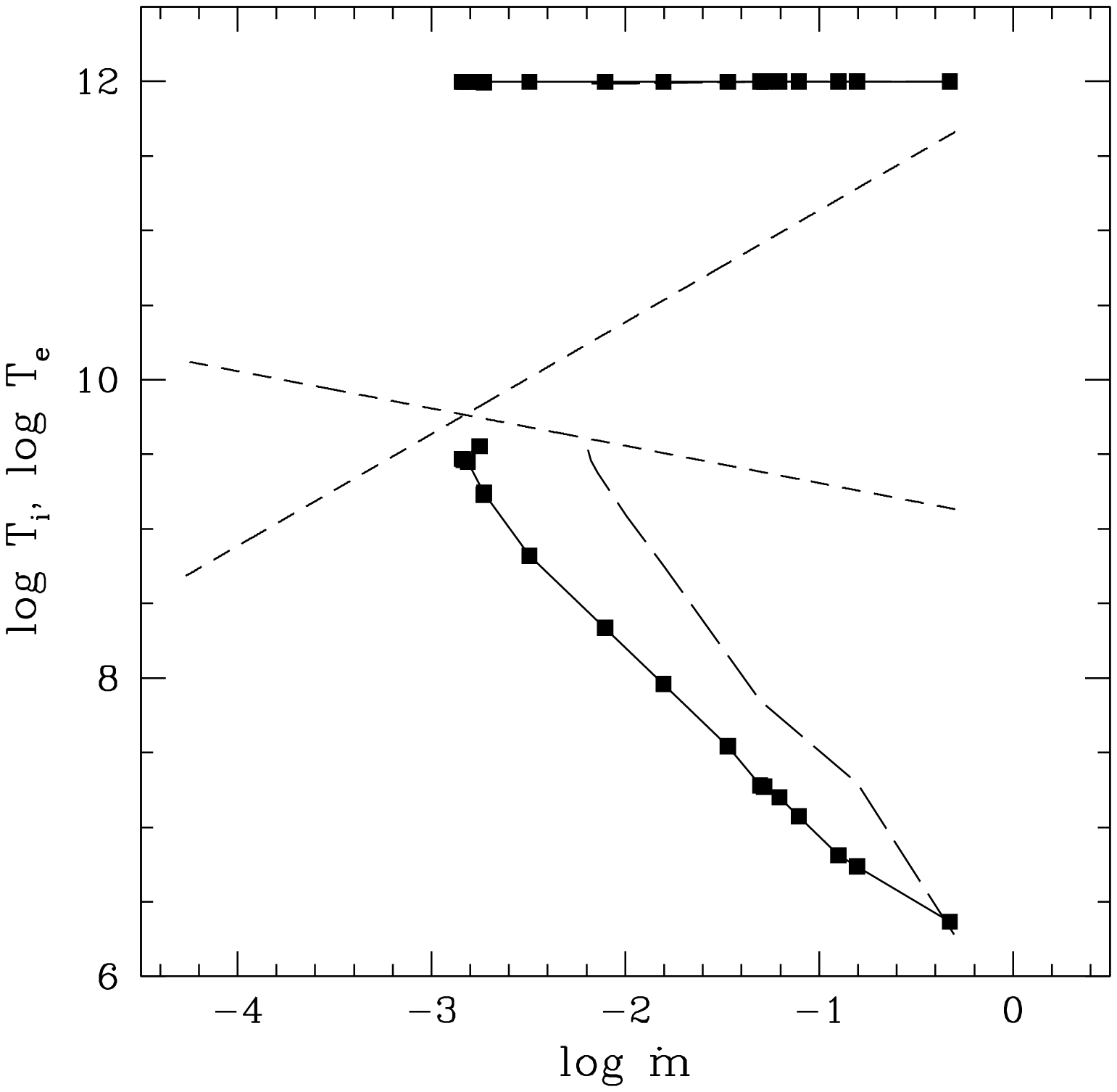}
\caption{The dependence of the ion and electron temperature at the top of the
transition zone on
the total accretion rate at $10 \RS$: equilibrium solution (continuous line
with squares), solution with evaporation/condensation (long dashed line)
and prediction from 
WC\. Z97 (short dashed line).}
\label{fig:Ti10}
\end{figure}

\begin{figure}
\epsfxsize=8.8cm \epsfbox{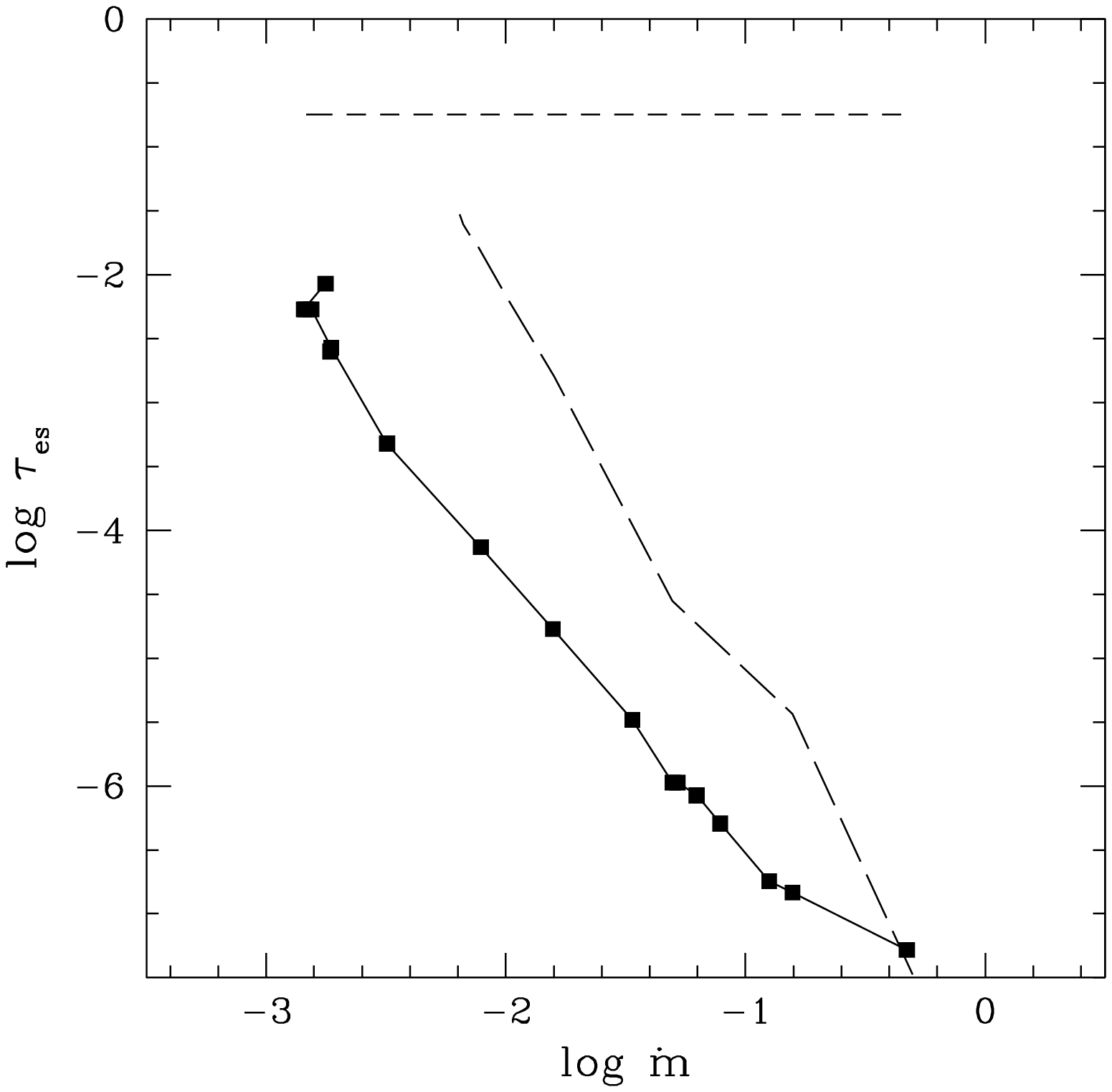}
\caption{The dependence of the optical depth of the corona on
the total accretion rate at $10 \RS$: equilibrium solution (continuous line
with squares), solution with evaporation/condensation (long dashed line)
and prediction from 
WC\. Z97 (short dashed line).}
\label{fig:tau10}
\end{figure}

Since the corona is not isothermal there is no single value of the ion and
electron temperature for a given radius. However, the representative values
are those at the top of the transition zone, at $\Pbase$. We plot them also
as functions of the accretion rate (Figs.~\ref{fig:Ti10}). We see that the trends are similar
to those determined by WC\. Z97 but there are quantitative 
differences. The ion temperature is always very close to the virial 
temperature and, at that radius, it never approaches the electron temperature. Finally, we plot the optical depth of the corona at the same
radius (Fig. \ref{fig:tau10}). The corona is much thinner than predicted 
by WC\. Z97, its optical depth strongly depends on the accretion rate.

\subsubsection{Radial trends}

All quantities depend not only on the accretion rate but also on the disk
radius. Therefore we study now the radial dependences of the coronal parameters
for a representative value of the accretion rate $\dot m=0.05$.

\begin{figure}
\epsfxsize=8.8cm \epsfbox{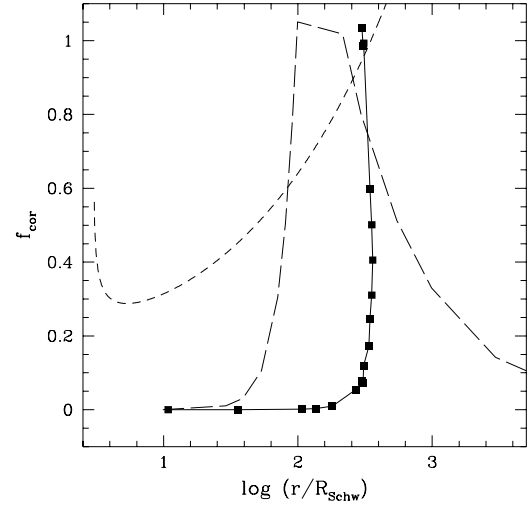}
\caption{The dependence of the fraction of energy generated in the corona on
the disk radius for the total accretion rate $\dot m=0.05$:
 equilibrium solution (continuous line
with squares), solution with evaporation/condensation (long dashed line)
and prediction from 
WC\. Z97 (short dashed line).}
\label{fig:fdiskr}
\end{figure}

\begin{figure}
\epsfxsize=8.8cm \epsfbox{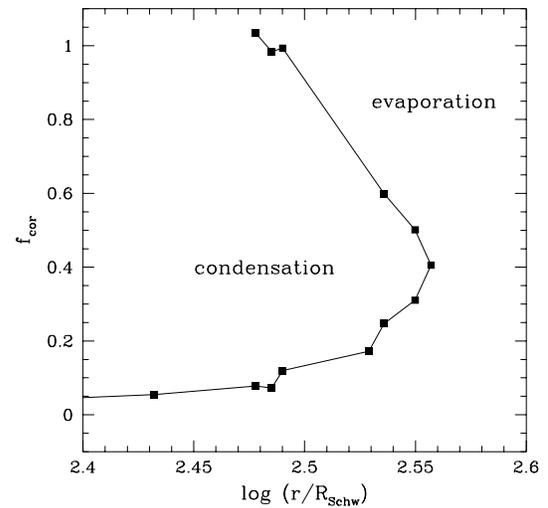}
\caption{The dependence of the fraction of energy generated in the corona on
the disk radius for the total accretion rate $\dot m=0.05$ - version
of Figure~\ref{fig:fdiskr} expanded near the outer edge of the coronal 
equilibrium solution. }
\label{fig:fdiskrb}
\end{figure}

There is an essential, qualitative change with respect to the results obtained
by WC\. Z97. 

In WC\. Z97
the strength of the corona increases outwards: corona is weak close the
black hole while all energy is generated within the corona at a certain
critical distance $\rmax$. Beyond this radius the disk/corona 
configuration does not exists. According to our new description of the 
disk/corona transition under no evaporation condition the solutions at
radii larger than a certain $\rmax$ also do not exists but at a radius only
marginally smaller than $\rmax$ the fraction of energy generated 
in the corona is not 1, but about 0.5 (Fig.
\ref{fig:fdiskr}). In a very narrow radial region below $\rmax$ 
two solutions exist: one with weaker corona and the other one with strong 
corona, which merge at $\rmax$. The strong corona solution reaches the value
$\fcor=1$ at a radius not much lower than $\rmax$ and the solution does not
continue down to smaller radii (mathematically, it exists as unphysical
solution with $\fcor>1$). Therefore this branch of solution does not  
offer a 
global solution for the corona. On the other hand the weak corona branch 
continues down to the marginally stable orbit and this branch represents the
solution acceptable under adopted conditions (no evaporation/condensation).
The nature of this multiple branch solutions is connected with the presence
of advection, as studied for the case of WC\. Z97 condition 
of disk/corona transition by Janiuk, \. Zycki \& Czerny (2000). The advection
dominated branch tends to vanish if the solution is computed more accurately by
determination of the coefficient $\delta$ iteratively from the radial
derivatives of physical quantities.

Therefore, according to our model, the corona in equilibrium with the disk
covers only a finite part of the disk, its strength increases outwards but 
it never dissipates more than about a half of the gravitational energy
available at a given radius (Fig.~\ref{fig:fdiskrb}). 
The transition from outer part without a coronal
solution to the inner part with corona is sharp  if studied within the
frame of solutions with no disk/corona mass exchange and the true nature of this
transition can be understood only within the frame of the solutions with
evaporation (see Section~\ref{sect:resevap}).

\begin{figure}
\epsfxsize=8.8cm \epsfbox{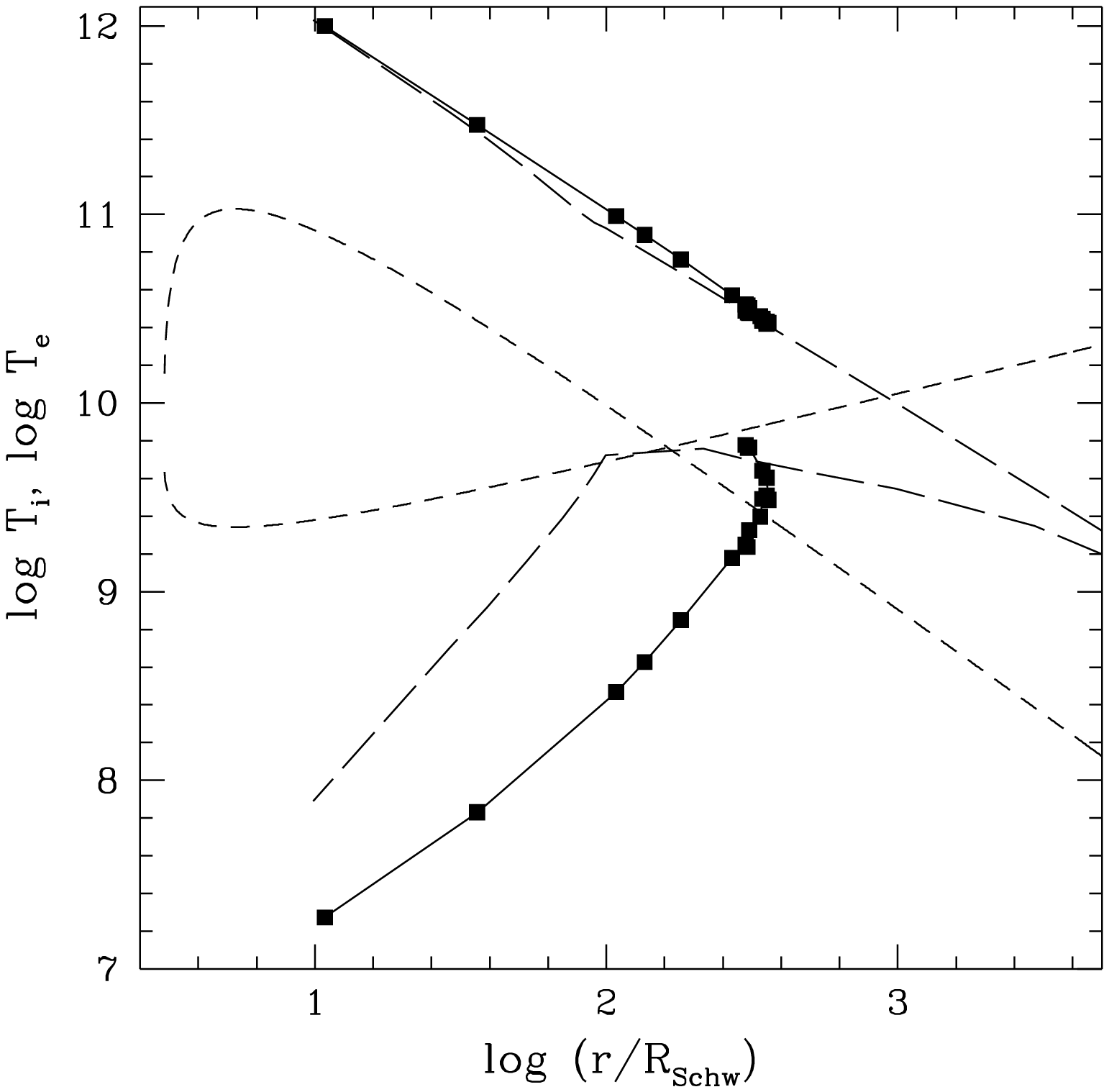}
\caption{The dependence of ion and electon temperature at the top of the
transition zone on
the disk radius for the total accretion rate $\dot m=0.05$: 
 equilibrium solution (continuous line
with squares), solution with evaporation/condensation (long dashed line)
and prediction from 
WC\. Z97 (short dashed line).}
\label{fig:Tier}
\end{figure}

The ion temperature at the top of the transition zone is always quite close 
to the local virial temperature so it clearly decreases with the radius
(Fig.~\ref{fig:Tier}).
On the other hand the electron temperature at $\Pbase$ increases outwards on
the basic branch, as in WC\. Z97. Its value at the outer edge of the
coronal solution is of the same order as predicted by WC\. Z97 
but closer to the central object is much cooler.

\begin{figure}
\epsfxsize=8.8cm \epsfbox{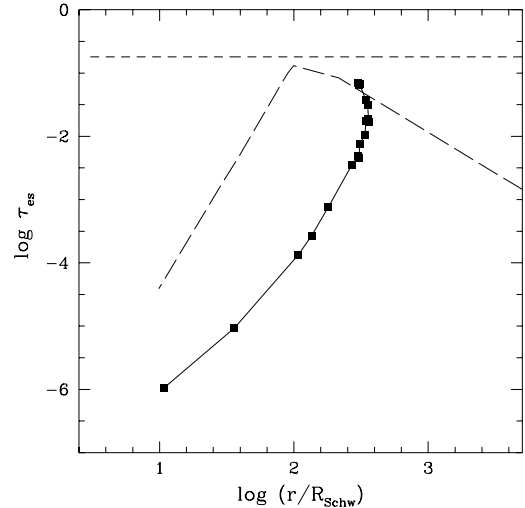}
\caption{The dependence of the optical depth of the corona on
the disk radius for the total accretion rate $\dot m=0.05$: 
 equilibrium solution (continuous line
with squares), solution with evaporation/condensation (long dashed line)
and prediction from 
WC\. Z97 (short dashed line).}
\label{fig:taur}
\end{figure}

The optical depth of the corona is not constant, as in WC\. Z97,
but it strongly decreases inwards along the basic branch
(Fig.~\ref{fig:taur}). Therefore, the
effect of Comptonization apart from the very outer edge of the coronal
solutions is completely negligible as long as the inner disk is not
evaporated (see Section~\ref{sect:resevap}).
In this paper we neglect Compton heating, since the accretion is a main
source of heating. However, Compton heating may increase  the optical
depth of the base of the corona and we plan to study this effect in
a future work.

The radial trends change with the accretion rate is a way similar to that
described by WC\. Z97. The extension of the corona, $\rmax$, 
is large for large accretion rate and decreases with  decreasing
accretion rate. For accretion rates below $\dot m=1.68 \times 10^{-4}$ 
no equilibrium solutions are
obtained. Again, the nature of this phenomenon can be understood by
analyzing the solutions which allow for mass exchange in the disk/corona 
system (see Section~\ref{sect:resevap}).

\subsection{Solutions with evaporation/condensation}
\label{sect:resevap}

Construction of those solutions is more complex as the solutions have global
character, i.e. the results at a given radius are coupled to those at other
radii through the continuity equation for the corona (Equation~\ref{eq:cont}). 
Since we follow semi-analytical approach to solution of Eq.~\ref{eq:cont} we
first present local numerical solutions which are parameterized by the coronal 
accretion rate, $\mcor$, in addition to global parameters $M$, $\dot m$ and 
$\alpha$, and we give the analytical formulae approximating the relation 
between $\mcor$ and $\mdotz$.

The vertical structure of the disk/corona system with evaporation is qualitatively 
similar to the case without mass exchange. However, the global trends
are essentially different so in this section we concentrate on those
aspects of the model.

\begin{figure}
\epsfxsize=8.8cm \epsfbox{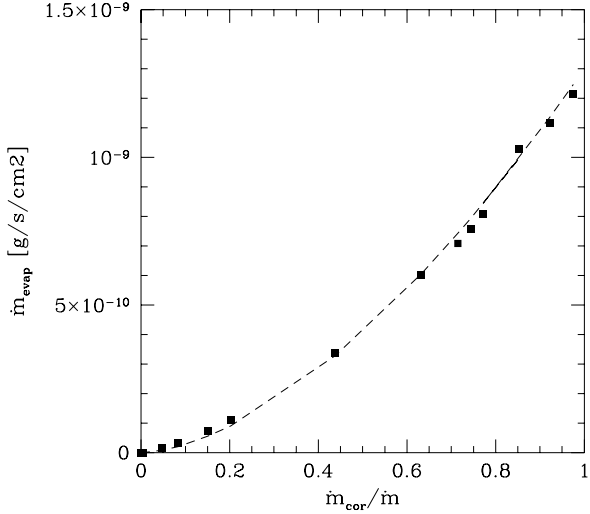}
\caption{The local relation between the accretion rate in the corona and the
evaporation rate of the disk at $10 \RS$  
for the total accretion rate $\dot m=5 \times 10^{-4}$. Dashed line shows 
the predictions from analytical formula \ref{eq:mdotz_loc_evap}.}
\label{fig:evap23}
\end{figure}

\begin{figure}
\epsfxsize=8.8cm \epsfbox{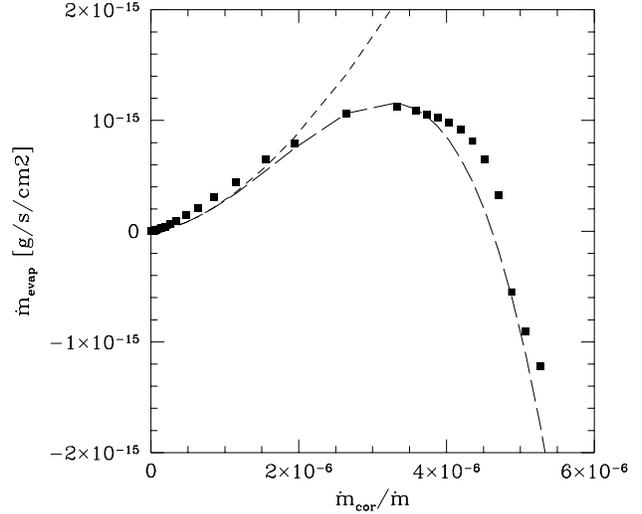}
\caption{The local relation between the accretion rate in the corona and the
evaporation rate of the disk at $10 \RS$  
for the total accretion rate $\dot m=0.05$. 
Short dashed line shows 
the predictions from analytical formula \ref{eq:mdotz_loc_evap} and long 
dashed line from \ref{eq:mdotz_loc_gen}.}
\label{fig:evap25}
\end{figure}

In Fig.~\ref{fig:evap23} we show the dependence of the evaporation 
rate of the cool
disk on the accretion rate in the corona at $10 \RS$ for a small 
value of the total
accretion rate ($\dot m= 5 \times 10^{-4}$). From Fig.~\ref{fig:fdisk10}
we see that we should not expect an equilibrium solution without evaporation
so it is not surprising that the evaporation rate $\mdotz$ is positive for
all values of the accretion rate within the corona from 0 to the total 
accretion flux. 

The relation is well approximated by an analytical formula
\begin{equation}
\mdotz= 1.27 \times 10^{-2} \mcor^{5/3} R^{-3/2} \alpha_{0.1}^{-7/2} 
M_{\rm BH8}^{-1}
\, [{\rm g s^{-1} cm^{-2}}].
\label{eq:mdotz_loc_evap} 
\end{equation}
Here, viscosity is described in dimensionless units: 
$\alpha_{0.1} = \alpha /0.1$.
This formula neglects the effect of condensation and it is a good 
approximation
of the solution only for very low accretion rates $\dot m$. Applicability
of this asymptotic expression will be qualitatively discussed at the end of
this Section.

Such an analytical expression can be now conveniently used to determine the
global solution by substituting it to the continuity equation \ref{eq:cont}.

The radial dependence of the coronal accretion rate and evaporation rate are
therefore:
\begin{equation}
{\mcor}= 3.09  {\alpha_{0.1}}^{21/4} R^{-3/4}
\label{eq:glob_evap} 
\end{equation}
\begin{equation} 
\mdotz = 8.23 \times 10^{-2} M_{\rm BH8}^{-1} \alpha_{0.1}^{21/4} R^{-11/4}.
\end{equation}

These formulae are qualitatively similar to the formulae derived by MMH94 
in the case of a single-temperature plasma and 
no condensation condition at the basis of the corona. The fraction of mass
carried by the corona increases inwards. Finally, at a certain radius 
$\Revap$ all
the mass is in the coronal flow due to the complete disk evaporation: $\dot m
= \mcor$ (Fig. \ref{fig:fdiskr} long dashed line). Its value
\begin{equation}
{\Revap} = 4.51 {\alpha_{0.1}}^{7} {\dot m}^{-4/3}
\label{eq:Revap} 
\end{equation}
is surprisingly similar to the value determined by Liu et al. (1999) on the
basis of Mayer \& 
Mayer-Hoffmeister's results. The only
essential difference is the dependence on the viscosity present in our 
approach but absent in theirs. This is basically the result of our different 
description of the conduction flux close to the disk surface and the
presence of the additional
step in energy transfer in the form of ion-electron coupling which results in
the electron temperature much lower than the virial temperature, depending 
also on the viscosity. 

Assuming the viscosity parameter 
$\alpha = 0.1$ we obtain practically the same results as Liu et al. (1999)
but larger (smaller) viscosity in our case result in larger (smaller)
purely 'coronal' region. The dependence on the accretion rate is
almost identical in both cases (see also Section~\ref{sect:transi}).

We now proceed to the case of large accretion rate. From 
Fig.~\ref{fig:fdisk10} we see that for large accretion 
rate $\dot m=0.05$ at 10 $\RS$  there is an 'static' equilibrium
solution so generally we can expect either evaporation, or condensation, 
depending on the value of the accretion rate within the corona itself.  
Fig.~\ref{fig:evap25} confirms those expectations. For extremely low values
of coronal accretion rate $\mdotz$ is still well approximated by the analytical
formula (Eq. \ref{eq:mdotz_loc_evap}) but a turnover appears at larger values and
finally evaporation is replaced with condensation, as $\mdotz$ becomes
negative.

Therefore we need to replace Equation \ref{eq:mdotz_loc_evap} with 
more general formula able to represent both evaporation and
condensation phenomenon. 
\begin{eqnarray}
\lefteqn {\mdotz  =  1.27 \times 10^{-2} {\mcor}^{5/3} 
R^{-3/2} {\alpha_{0.1}}^{-7/2} M_{\rm BH8}^{-1} \times \left[1- \right.} 
\nonumber \\ 
 & & \left. -1 \times 10^{42} {\dot m}^{10}{\mcor}^{5/2} 
  R^{-12.5} {\alpha_{0.1}}^{-8.5} \right ]  \, 
[{\rm g s^{-1} cm^{-2}}].
\label{eq:mdotz_loc_gen}  
\end{eqnarray}

\begin{figure}
\epsfxsize=8.8cm \epsfbox{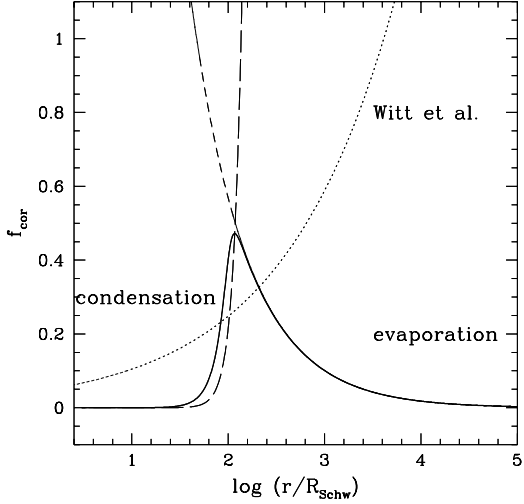}
\caption{The global solution for the radial dependence of the fraction of 
energy generated in the corona for the total
 accretion rate $\dot m=0.05$ and $\alpha = 0.08$ 
(thick continuous line). Short dashed line shows 
the predictions from analytical formula (\ref{eq:glob_evap}), long 
dashed line from an analytical approximation to numerical results shown 
in Fig.~\ref{fig:fdiskr} as a continuous line with squares, and dotted line shows the solution
of WC\. Z97.}
\label{fig:glob25}
\end{figure}

This formula is oversimplified as it does not take into account the 
non-monotonic two-value dependence of $\fcor$ on the radius for models in
equilibrium, for which $\mdotz=0$ (see Fig.~\ref{fig:fdiskrb}). However, it 
reproduces accurately enough the extension of the strong condensation zone.
More accurate description would require numerical computations of the vertical
structure of the corona together with integration of Eq.~\ref{eq:cont}.

Previously found global solution based solely on evaporation applies 
everywhere in the disk down to the marginally stable orbit if the second term in 
Eq.~\ref{eq:mdotz_loc_gen} is smaller than 1, i.e. when the total accretion rate
is below the limiting value 
$\dot m = 1.31 \times 10^{-3} \alpha _{0.1}^{17/25}$ for $R=3$. 
In objects with such a low accretion rate the corona formation proceeds
monotonically, corona carry an increasing fraction of mass as the flow 
approaches a black hole and finally, at radius given by Eq.~\ref{eq:Revap}
the disk disappears, as it is replaced by an optically thin flow of ADAF
type. 

In objects with significantly higher accretion rate 
the situation is more complex
than in low accretion rate case. The equations ~\ref{eq:mdotz_loc_gen} 
and \ref{eq:cont} have to be
now solved numerically. An example is shown in Fig.~\ref{fig:glob25}
(thick continuous line).
At the outermost part of the disk the corona formation proceeds according
to Equation~\ref{eq:glob_evap}, as before. Closer in the cooling becomes
more efficient and the corona formation proceeds more slowly than predicted
by Equation~\ref{eq:glob_evap}. When we approach the radius for which an
equilibrium solution exists (i.e. possible balance between the disk and the
corona without evaporation, see Sect.~\ref{sect:noevap}) the corona reaches its maximum
strength. In the innermost part of the disk the corona strength decreases
as the evaporation is replaced by condensation. The radial dependence of the 
fraction of energy 
generated in the corona in this region 
is qualitatively similar to the result given under
assumption of no mass exchange between the disk and the corona. Trends are
the same as those given by WC\. Z97 although the actual radial
dependence is much steeper.

\begin{figure}
\epsfxsize=8.8cm \epsfbox{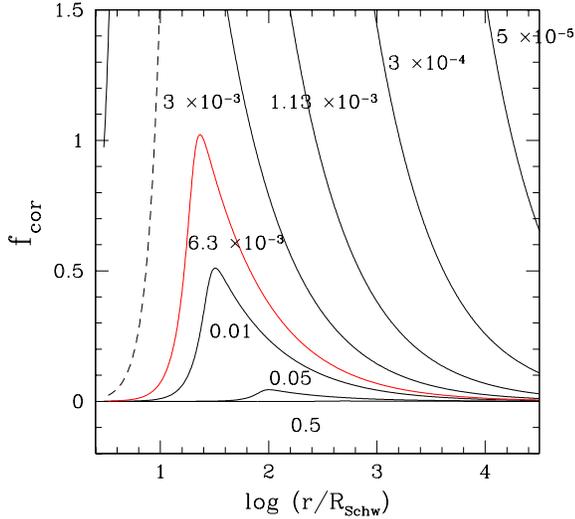}
\caption{The global solution for the radial dependence of the fraction of 
energy generated in the corona for $\alpha = 0.05$ and the accretion rates
equal $0.5$, $0.05$, $6.3 \times 10^{-3}$, $3\times 10^{-3}$, 
$1.13 \times 10^{-3}$, $3 \times 10^{-4}$, and $5 \times 10^{-5}$.  
Dashed line shows the formal continuation of the solution
for accretion rate $3 \times 10^{-3}$ - reconstruction of the disk due to 
condensation.}
\label{fig:globseq}
\end{figure}

A sequence of solutions showing the radial dependence of the fraction of
the energy generated in the corona for several values of the accretion rate is
shown in Fig.~\ref{fig:globseq}.
For significantly higher accretion rates the maximum strength of corona 
never approaches the value of 1, so the two-phase flow extends to the
marginally stable orbit.
When the total accretion rate is intermediate the evaporation is efficient 
enough
to cause a transition to an optically thin flow ($\fcor=1$) 
at a radius approximately
given by Eq.~\ref{eq:Revap}. However, in the innermost region 
the condensation is
efficient if there is large soft photon flux available to cool electrons
through Comptonization. We expect that it might lead to cool clump 
formation or even a reconstruction of the disk in this region. 
Such a secondary disk rebuilding is now present in our model for accretion
rates higher than $\dot m^{\rm min}$.
More accurate numerical
results for an accretion rate which leads to an ADAF type flow in the
innermost part give the value
\begin{equation} 
{\dot m^{\rm min} } = 1.9 \times 10^{-3} \alpha_{0.1}^{17/25}.
\label{eq:mdotmin}
\end{equation}

However, the current version of our model
does not allow to describe yet the details of the transition from 
optically thick disk
through optically thin disk to complete disappearance of the disk flow
and eventual reverse of this process in the innermost part of the flow.
Therefore, at present, we usually assume that once the transition occurs from 
the 
flow from disk/corona geometry to an optically thin flow at a certain 
radius it means that
there is no disk flow closer to a black hole. However, we also discuss an
opposite case in Section~\ref{sect:mdotsig}. 

Further decrease
of the accretion rate results in a smooth expansion of the ADAF zone. 
In all other models the first appearance of the single phase hot flow
is expected at the marginally stable orbit (e.g. Esin et al. 1997). 
In our model
the disk either extends to the marginally stable orbit or down to a radius
larger than the minimum radius dependent on the viscosity, with no solutions
for the transition radius ever expected between these two radii. This property
of our solutions is caused by the domination of the condensation over 
evaporation in the innermost part of the disk where the pressure is large.

\begin{figure}
\epsfxsize=8.8cm \epsfbox{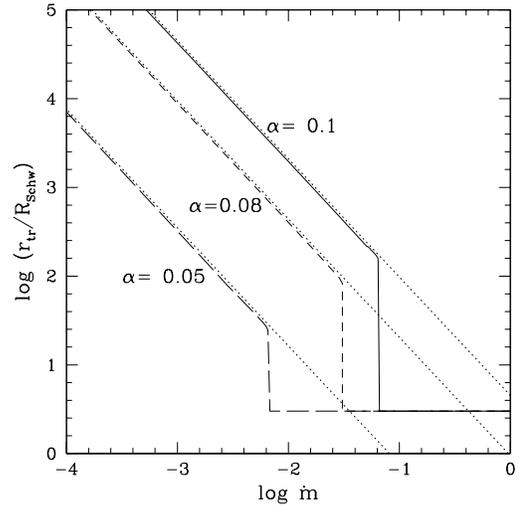}
\caption{The dependence of the transition radius from disk/corona flow to
an ADAF flow on the accretion rate for the viscosity parameter $\alpha = 0.1$
(continuous line), $\alpha=0.08$ (dashed line) and $\alpha=0.05$ 
(long, dashed line). Dotted lines show the
corresponding predictions based on the pure evaporation formula 
(Eq.~\ref{eq:Revap}).}
\label{fig:rtran}
\end{figure}

The dependence of the transition radius on the accretion rate
is shown in Fig.~\ref{fig:rtran}. We see that for low accretion rates the 
position of the transition radius depends on the accretion rate and the 
relation is well described by Eq.~\ref{eq:Revap}, as expected. However, 
at a certain value of the accretion rate the condensation takes over and 
the coronal dissipation decreases in the innermost part of the disk enough 
to prevent the transition to an ADAF flow. This critical accretion 
rate $\mdotevapbranch$ is well described by the two analytical 
conditions: Eq.~\ref{eq:Revap} and $\mdotz=0$ 
(see Eq.~\ref{eq:mdotz_loc_gen}) which leads to
\begin{equation}
\mdotevapbranch = 6.92 \times 10^{-2} \alpha_{0.1}^{3.3}.
\label{eq:evapbranch}
\end{equation} 

This accretion marks the formation of the evaporation branch of the 
disk/corona solution which is essential for explanation of specific 
luminosity states of accreting black holes and time-dependent behaviour 
of X-ray sources (see also Sec.~\ref{sect:mdotsig}).

However, the essential point which directly
results from radial computations is the understanding of the inner 
region without
disk/corona solutions. What WC\.Z97  interpreted as a 'bare' disk is
actually a 'bare corona', or ADAF type of flow.

Determination of the coronal accretion rate as a function of the total
accretion rate is a starting point of the discussion of the stability of 
solution and eventual time-dependent evolution.

\subsection{Expected time-dependent behaviour}
\label{sect:mdotsig}

\subsubsection{Surface density of the accretion flow}

The stability and time evolution of the accretion flow is most conveniently
discussed on the plot showing the accretion rate, $\dot m$, 
and the surface density
of the flow, $\Sigma$, at a representative radius.

We calculate the fraction of the energy generated in the corona at by
numerical integration of Eq.~\ref{eq:cont}, assuming the analytical formula 
for the local evaporation/condensation rate given by 
Eq.~\ref{eq:mdotz_loc_gen}. 
Knowing the corona
properties we solve the cold disk structure at this radius, with the 
appropriate boundary conditions (see Sec.~\ref{sect:cold}).

\begin{figure}
\epsfxsize=8.8cm \epsfbox{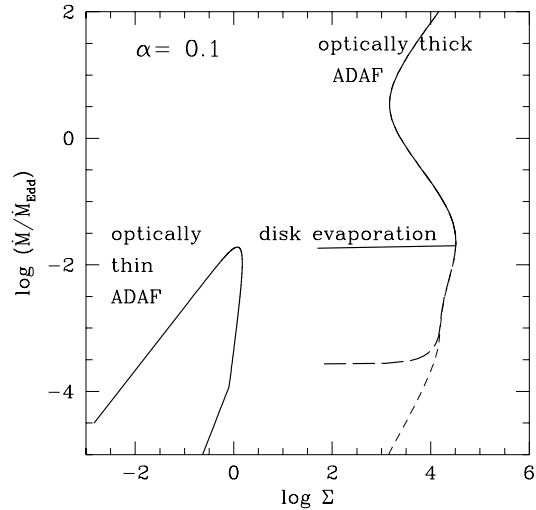}
\caption{The relation between the accretion rate and the surface density
of the disk/corona system at $10 \RS$ for the viscosity parameter 
$\alpha = 0.1$ in case
of irreversible  ADAF transition
(continuous line) and with secondary disk rebuilding (long dashed line).
Short-dashed line shows the standard Shakura-Sunyaev model supplemented
with advection essential at high accretion rates, calculated for $M=10^8 
M_{\odot}$ - slight wiggles of the gas dominated lower branch are caused 
by bound-free opacities.  
Continuous line in the lower right part of the diagram shows solution for 
an optically thin flow (Narayan \& Yi 1995b, 
Abramowicz et al. 1995 ),  computed using the
code of  Zdziarski (1998).}
\label{fig:mdotsig}
\end{figure}

Generally, for $\dot m > \dot m _{\rm evap.branch} $ the 
strength of corona is small and the 
solution is practically the same like in the thin disc without corona.
However, when $\dot m$ approaches $\dot m_{\rm evap.branch}$
this is no longer true and the solution starts to depart  horizontally 
from standard $\log \dot m$--$ \log \Sigma$ curve (Fig.~\ref{fig:mdotsig}).
The plot depends on a distance from the black hole, i.e. whether
it is higher or lower than the critical radius for which the curve  
in Fig.~\ref{fig:globseq} has maximum at $f_{\rm cor}=1$. 
The critical radius can be determined analytically by combining 
Eq.~\ref{eq:Revap} and Eq.~\ref{eq:evapbranch} which gives
\begin{equation}
\Rec10 = 158.7 \alpha_{0.1}^{2.6}.
\end{equation}

For $R>\Rec10$, $f_{\rm cor}$ increases with decreasing $\dot m$, and 
finally $f_{\rm cor} =1$ for a certain value of accretion rate
$\dot m_{\rm evap}$ smaller then  $m _{\rm evap.branch} $. As 
$f_{\rm cor} \rightarrow 1$, the disc surface density drops with 
$\dot m \rightarrow \dot m _{\rm evap}$.

For $R< \Rec10$ the plot depends on the assumption of considering or
rejecting the solutions representing the secondary disc 
condensation (see dashed line in Fig.~\ref{fig:globseq} and 
Sec.~\ref{sect:resevap}). 
In  Fig.~\ref{fig:mdotsig} we show  the relation between the  accretion
rate and the  surface density for $r=10R_{\rm Schw}$ and $\alpha=0.1$. 

If we assume that once the disk is evaporated it
never forms again closer to the black hole, then the  transition radius 
between the disk/corona and single-phase ADAF, suddenly drops to the 
marginally stable orbit.  In such case, when decreasing accretion
rate attains the value of $m _{\rm evap.branch} $, 
the discontinuous jump in the corona strength occurs from $f_{\rm cor} <1$  to 
$f_{\rm cor} = 1$. 
This sharp transition is marked as a horizontal line called evaporation branch.
In time-dependent solution the disc structure may actually follow this branch
if $\dot m$ decreases suddenly. 

On the other hand, if we allow for existence of the reconstructed branch of 
the disk at the innermost part of the disk, whenever it is allowed, the 
solution enters the evaporation branch gradually, it is 
shallow and positioned at much lower accretion rate (long dashed line). 
However, such a solution has 'an ADAF hole' for a certain range of 
radii. In further considerations we adopt the first
solution as the correct one since the inner secondary disk formation is
not consistently described within the frame of our model and may not
represent a physically acceptable solution.

The mechanism of disk evaporation leads to departure from standard optically 
thick solution in the form of practically horizontal branch which develops
at intermediate accretion rates. 
For $\alpha=0.1$ the position of this branch is
approximately at the same value of the accretion rate as the maximum of
the accretion rate at the optically thin branch,  as described by 
Narayan \& Yi (1995b), Abramowicz et al. (1995) and Chen et al. (1995),
but this is not the case for other values of $\alpha$ since the
optically thin branch was computed for a fixed assumed value of the
Compton parameter characterizing the cooling of the optically thin 
flow. Correct solution should be obtained taking into account the
actual soft X-ray flux coming from outer parts of the disk and should
be a natural extension of the two-phase solution in the case of cold
disk absence. At present we
cannot extend our horizontal branch to the low surface density range and to
reproduce the solution for a single phase medium 
since the present model is based on
assumption that the underlying disk is optically thick and absorbs half of
the X-ray radiation emitted by the corona at a given radius and reemits this
radiation in the form of soft flux. The description of the transition
from a two-phase flow to a single-phase optically thin flow would require
relaxation of this assumption and a number of other modifications, like 
non-local description of the soft photon radiation field since the outermost
part of the ADAF flow will be predominantly cooled by soft photons emitted by 
the outer parts (e.g. Esin et al. 1997) and/or soft photons originating from
synchrotron radiation. We plan to study this problem in the 
future, in order to give a complete view of the transition between those two
branches. 

The position of evaporation branch in our model depends on the 
viscosity - lower viscosity parameter $\alpha$ corresponds to lower
accretion rate at this branch. For $\alpha = 0.1$ and large 
black hole mass it joins the (almost) standard Shakura-Sunyaev branch
just below the turning point marking the gas pressure dominated branch.
Therefore, the transition practically happens from unstable radiation
pressure dominated branch to evaporation branch with a very narrow
range of accretion rates in between. Adopting lower value of viscosity,
however, broaden this range considerably. For low mass of the black hole the
stable gas dominated branch exists even for $\alpha = 0.1$ since the 
position of the transition from radiation pressure to gas pressure 
dominated solution in Shakura-Sunyaev disk depends on the mass of the c
entral object.

As seen from Fig.~\ref{fig:rtran}, the position of the evaporation branch 
does not depend on radius in the innermost part of the disk. 
At $R<\Rec10$ the position of the evaporation branch is given 
by Eq.~\ref{eq:evapbranch} while for $R > \Rec10$ this position is 
determined by Eq.~\ref{eq:Revap}. This property has important 
consequences for the limit cycle evolution expected for a range 
of accretion rates corresponding to the unstable radiation pressure 
dominated branch of thick disk solution.

\subsubsection{Stationary solutions and outbursts}

This diagram (Fig.~\ref{fig:mdotsig}), although not complete, 
shows the importance of the evaporation branch for stationary solutions 
and time-dependent evolution of accretion disk.
If the accretion rate is larger than the Eddington limit the model is located
on the stable optically thick ADAF branch. Stationary solution well represents
an accretion process at this parameter range. 
When the accretion rate is very low, below
the evaporation branch, the flow is described by the optically thin 
stable ADAF solution in the innermost part, and a disk/corona flow in the 
outer part. The solution should be also well represented by a stationary model,unless the disk/corona interaction lead to some instabilities.  

However, for intermediate accretion rates, the disk/corona flow, as the
classical radiation pressure dominated model, may
display a complex limit cycle, as already studied, without taking into account
 the evaporation branch, by Szuszkiewicz \& Miller (1998), Honma, Matsumoto \&
Kato (1991) and
Janiuk et al. (2000). The solution would oscillate between the upper branch
during the eruption and 
and the evaporation branch between the outbursts. Actual position of the upper 
branch may be at lower accretion rates, if the strong outflow in addition to
radial advection, is included (e.g. Nayakshin, Rappaport \& Melia 1999, 
Janiuk et al. 2000).
The timescale for such evolution is shorter than the viscous timescale of the
standard Shakura-Sunyaev stationary disk at this radius because of the reduced
surface density. The exact prediction of the global outburst involves the
entire unstable part of the disk so, without actual computations, the
exact timescale and the amplitude cannot be determined but order of magnitude
timescales involved are seconds for galactic sources and months/years for
AGN. 

The character of this evolution would depend significantly on the radial
extension of the instability zone. If this extension is below $\Revac$ 
the position of the evaporation branch is the same for all radii and the 
entire disk will oscillate between the upper (advection or outflow) 
dominated branch and the evaporation branch so the optically thick 
disk will be present during the entire cycle. However, if the radial 
extension is larger than $\Revac$ the evaporation branch at the outermost 
radius involved in the outburst is lower than evaporation branch inside. 
The accretion rate in low luminosity phase of the outburst is then too 
low for optically thick disk in the innermost part, the transition to ADAF 
takes place so the optically thick disk will be temporarily evacuated 
during such a cycle. It is interesting to note that probably both kinds 
of outbursts are seen in the case of microquasar GRS 1915+105, as two 
distinct kind of spectral behaviour and of the relation between the 
length of outburst and the frequency of QPO phenomenon are 
observed (Trudolyubov, Churazov \& Gilfanov 1999).
Also the observed X-ray delay in the outburst of soft X-ray transient 
GRO J1655-40  shows that most probably the system made the transition from 
an inner ADAF/outer disc/corona state to a pure disc/corona state 
Hameury et al. (1997).
The same transition is seen in Cyg X-1 (Esin et al. 1997).

The expected  amplitude of the outburst is considerably reduced by the 
presence of the evaporation branch - expected luminosity changes are by
an order of magnitude while, if the gas pressure dominated branch is the lower
branch of the limit cycle the outburst of up to four orders of magnitude
would be expected, against any observational evidence for the timescales
involved.

According to our model, there are strong systematic differences expected 
between the AGN and GBH since the range of the radiation pressure instability 
depends on the central mass while the position of the evaporation branch does 
not. If the evaporation branch is positioned significantly
below the radiation pressure dominated branch we can also expect a stable 
accretion flow in the accretion rate range corresponding to the gas pressure 
dominated solution. 
It means that, independent on the viscosity, the range of accretion rates
corresponding to stable, gas dominated solution is always broader for 
galactic black holes than for AGN. In particular. no such range at all 
is predicted in AGN for $\alpha=0.1$ since this solution is entirely 
replaced by evaporation branch and ADAF. 
 
\section{Discussion}
\label{sec:dis}

\subsection{Comparison to other models}

\subsubsection{Underlying assumptions}

Our approach to modeling the disk/corona structure follows the MMH94 attitude.
We also model the energy dissipation within a corona as due to the accretion 
flow, we take the transition between the disk and the corona which allows to
determine the fraction of energy generated in the corona as a function of 
radius as well as the transition radius to the optically thin flow as a 
function of the mass of the black hole, accretion rate and the viscosity 
parameter in the corona. 

There are, however, major important differences between our model and MMH94:
(i) our corona is basically a two-temperature medium with energy dissipation
due to ion collisions and the Coulomb transfer of energy from ions to 
electrons, as in SLE and ADAF models (ii) apart from bremsstrahlung we allow
also for Compton cooling (iii) we determine the conductive flux at the basis
of the corona for the adopted heating/cooling mechanism instead of using the
scaled prediction from solar corona. This last modification is responsible for
the dependence of the accretion rate in the corona on the viscosity parameter
in the case of our model and the lack of such dependence in MMH94 and
subsequent papers (Liu, Meyer \& Meyer-Hofmeister 1995, Liu et al. 1999, 
Meyer, Liu \& Meyer-Hofmeister 2000). The 
predicted dependence of the coronal accretion rate on the radius 
is qualitatively similar in both papers in the
case of low accretion rate but not in the case of large accretion rate when the
condensation, neglected by MMH94, is important. Our approach to conduction is
therefore the same as of Dullemond (1999) but he did not consider a 
two-temperature plasma and Compton cooling.

Two-temperature accreting corona was discussed by WC\. Z97 but their condition
of disk/corona transition based on analogy with Compton heating medium was
not accurate numerically and missed an important insight on corona formation
in the innermost and outermost  parts of the flow.

\subsubsection{Transition radius to ADAF-type flow}
\label{sect:transi}

 The transition from an outer optically thick disk flow to an inner optically thin ADAF was predicted by a number of models. In some models the transition was
based on ADAF principle (i.e. the transition happens whenever an ADAF solution is possible), but in other cases the necessity of the transition resulted
from the model itself (MMH94 and applications, our model). It is therefore
interesting to compare quantitatively those estimates as they may in principle
be estimated observationally.

Condition based on the strong ADAF principle, i.e. on the absence of ADAF 
solutions:

\begin{equation} 
\rtr = 1.9 \times 10^4 \dot m^{-2} \alpha_{0.1}^4 \RS  
\end{equation}
(Abramowicz et al. 1995, Honma 1996, Kato \& Nakamura 1998).

Approximation to the results of Esin et al. (1997) for $\alpha = 0.25$ 
and $\beta = 0.5$:
\begin{equation}
\renewcommand{\arraystretch}{1.5}
\begin{array}{l@{\hspace{1cm}}c}
\rtr  = \times 10^4 \RS  &   \dot m  <   0.084   \\
\rtr  =  30 \RS  & \dot m  \approx  0.084   \\
\rtr  =  3 - 30 \RS & 0.084 <  \dot m  < 0.092  \\
\rtr  =  3 \RS  & \dot m  >  0.092
\end{array}
\renewcommand{\arraystretch}{1}
\end{equation}

Liu et al. (1999) gives:

\begin{equation} 
\rtr =18.3 \dot m^{-1/1.17}  \RS
\end{equation}

Our results:
\begin{equation}
\renewcommand{\arraystretch}{1.5}
\begin{array}{l@{\hspace{0.5cm}}l} 
\rtr  =  4.51 \dot m^{-1.33} \alpha_{0.1}^7 \RS & \dot m < 6.92 
\times 10^{-2} \alpha_{0.1}^{3.3}  \\
\rtr  =  3 \RS  &\dot m >  6.92\times 10^{-2}\alpha_{0.1}^{3.3} 
\end{array}
\renewcommand{\arraystretch}{1}
\label{por:tran}
\end{equation}

All those conditions are 
roughly in agreement with observational requirement that transition
between the high
and low state happens at  luminosity $10^{37} $ erg s$^{-1}$ (Tanaka 1999).
The accurate dependence of the transition radius on the accretion rate can in
principle be determined from the shape of the reflected component 
(Done \& \. Zycki 1999) in X-ray novae but the errors are considerable 
and the presence of the hot Compton heated skin on the disk surface 
further complicates the interpretation of the data. However, high 
precision spectroscopy from new X-ray satellites should give stronger 
constraints on the available models.

\begin{figure}
\epsfxsize=8.8cm \epsfbox{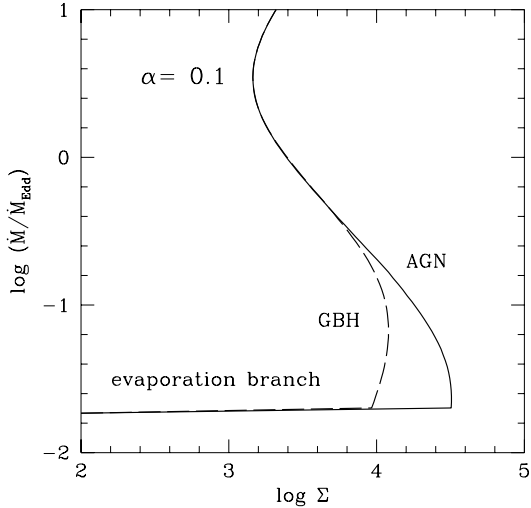}
\caption{The relation between the accretion rate and the surface density
of the disk/corona system at $10 \RS$ for the viscosity parameter $\alpha = 0.1$ in case
of irreversible  ADAF transition
for an AGN ($M= 10^8_{\odot}$, continuous line) and a GBH ($M= 10_{\odot}$,
dashed line). Stable solutions (positive slope of the curve) are expected for intermediate accretion rates
in the case of a small mass of the black hole but this region is very narrow
for massive black holes.}
\label{fig:mdotsig2}
\end{figure}

\subsection{Observational consequences}

\subsubsection{Luminosity states}

The transition from optically thick flow to ADAF, i.e. the accretion
rate at the evaporation branch marks the transition between the
hard state and soft state in accreting binaries, according to a
number of suggestions (e.g. Narayan 1996, Esin et al. 1997, 
Zdziarski, Lubi\' nski \& Smith 1999).   
When the accretion rate is high enough and two-phase flow exists up to the 
marginally stable orbit then the soft flux is always dominant, cooling of
corona is efficient and produces soft X-ray spectrum. Such a solution is stable in GBH 
for moderate accretion rates and may
correspond to the soft (high) state. If the accretion rate is very high the disk is 
unstable and evolving between the evaporation branch and advection-dominated 
optically thick branch - such a parameter range may be appropriate for a Very High 
State, with its complex spectral behaviour (e.g. Takizawa et al.
1997, Revnivtsev et al. 2000). 
On the other hand, when the accretion rate is very low, the innermost part 
of the disk is an optically thin ADAF, dominating the spectrum, so such a case
corresponds to the Low (or Hard) State. 

In our model, the position of horizontal evaporation branch does not depends 
on the mass of the black hole (Fig. \ref{fig:mdotsig2}). Nevertheless the
turning point between the unstable radiation pressure branch and 
the stable  evaporation branch is different for AGNs and GBHs and the possibility of formation of the High State in AGN is strongly suppressed.

Since most of the energy generates at about 10 $\RS$ we can roughly identify the
accretion rate ranges for various luminosity states on the basis of Fig~\ref{fig:mdotsig2}. Assuming $\alpha=0.1$ we obtain:
\begin{equation}
\renewcommand{\arraystretch}{1.5}
\begin{array}{l@{\hspace{1cm}}cc}
State & GBH  &  AGN   \\
Very~ High ~State  & \dot m  > 0.1  &  \dot m  > 0.07  \\
High ~State &   0.07 <  \dot m  < 0.1  & \\
Low ~State   & \dot m  <  0.07 & \dot m  <  0.07
\end{array}
\renewcommand{\arraystretch}{1}
\end{equation}

However, we stress that those limits depend significantly on the adopted value of the viscosity parameter $\alpha$.

\subsubsection{The nature of the Intermediate State in GBH}
  
The transition of X-ray novae from soft state with the dominant disk emission to a hard state 
with the dominant thin plasma (e.g. ADAF) contribution proceeds through the
Intermediate State. The analysis of the Nova Musca revealed that in the
intermediate state the reflection component is seen with an amplitude of 
order of 1 while the kinematical determination of the inner disk radius from
Doppler shifts of the reflected component show disk disruption (\. Zycki, 
Done \& Smith 1998). We can 
interpret such a state within the frame of our model as corresponding to 
an accretion rate for which a strong corona forms at intermediate radii. In 
this case the reflection is strongly biased towards larger radii instead of
towards innermost disk radius. Also very high values of the ionization 
parameter may be consistent with the irradiating source being directly above 
an accretion disk. Our interpretation is therefore different from that 
of Esin et al. (1997) since in their approach the intermediate state 
corresponds to
a disrupted disk with an innermost ADAF flow instead of the two-phase
disk/corona flow.  

\subsubsection{Variability}

The lack of High State in AGN, as predicted by our model for the viscosity
parameter $\alpha = 0.1$  
may offer an explanation why the observed variability
properties of AGN and GBH are not the same. GBH, when in their High 
State, dominated by disk emission, are only weakly variable in X-rays in the timescales of about a second, with rms 
amplitude of order of a percent (e.g. Takizawa et al. 1997).
Spectrally, such states are basically similar to quasar spectra. However,
quasars are strongly variable, with rms amplitude in the optical band dominated
by the disk emission is of order of 15 per cent (e.g. Giveon et al. 1999) in the 
timescales of years, so the simple scaling with the mass of the black hole does
not seem to hold. It may mean that quasars are actually counterparts of the
GBH in their Very High State. Since the spectra of GBH in Very High State are 
quite complex and not always dominated by the disk emission we can also expect
such weak disk spectral states among the quasar/Seyfert class. Well studies
source MCG-6-15-30 (Tanaka et al. 1995 and subsequent papers) may be an 
example, due to its exceptionally
low mass and consequently high luminosity to the Eddington luminosity ratio
(Nowak \& Chiang 1999).  

\subsubsection{Broad Line Region and LINERs}

Nicastro (2000) proposed a very interesting model of the Broad Line Region 
based on the disk/corona model of WC\. Z97. The basic idea is that the strong 
outflow develops at radii where the corona strength reaches its maximum and
cool clouds forming in this disk wind are responsible for the emission of 
broad lines. Predictions based on our new model does not alter presented 
estimates significantly. The radius of the maximum corona strength in the 
new model can be
determined from the condition $\mdotz = 0$ (see Eq.~\ref{eq:mdotz_loc_gen}),
using the result with negligible evaporation to determine the local coronal 
accretion rate (Eq.~\ref{eq:glob_evap}) which gives the dependence
\begin{equation}
R_{\rm max}=1.02 \times 10^3 \dot m^{16/23} \alpha_{0.1}^{37/115},
\end{equation} 
smaller by a factor of few than the WC\. Z97 formula (see Eq. 3 of Nicastro 
2000). On the other hand the
new corona efficiency is now roughly symmetric around this point so the 
weighted formula for the outflow used by Nicastro would not decrease this 
radius, as it did in the case of sharp edge corona of WC\. Z97 so the net 
effect will be
roughly the same as before.

It was suggested by Lasota et al. (1996), that LINERs may represent 
the systems with ADAFs.  
The transition from objects with broad emission lines to LINERs happens, 
according to Nicastro (2000), due to the switch off the outflow if the
accretion rate is too low for the existence of the radiation pressure 
dominated zone. The most efficient part of the corona, however, is located 
in the gas pressure dominated 
region so this connection is not clear. On the other hand, for very small
accretion rate, within the frame of our new model, there is a transition from
secondary disk rebuilding phase to a definite ADAF flow in the innermost part
(see Eq.~\ref{eq:mdotmin}).
Such a change may be important from the point of view of the radiative cooling,
the efficiency of advection and, consequently, the mass outflow.  
The mass loss through the wind as a function of radius 
is not calculated yet in the present model but 
this can be done in the future, following the method of e.g. 
Meyer \& Hoffmeister (1994) and WC\. Z97.

\begin{acknowledgements}
            
We are grateful to Drs R. Narayan and A. Esin, for their helpful comments
and suggestions on the original version of our paper.  
We thank A. Zdziarski for his computer program generating optically thin 
solutions used in Fig.~\ref{fig:mdotsig} and to Piotr \. Zycki for helpful
discussions.
Part of this work was supported by grant 2P03D01816 of the Polish
State Committee for Scientific Research.
\end{acknowledgements}

\appendix
\section{Evaporation/condensation rate at the basis of the corona}

The electron temperature profile within the transition zone is determined
by the effect of heating, radiative cooling, conduction and the vertical 
motion of the material (e.g. Begelman \& McKee 1990, McKee \& Begelman 1990, 
MMH94). 
The general form of this equation in a plane parallel geometry is the following
\begin{equation}
{d \over dz} \left[ \rho v_{\rm z} {5 \over 2} {k \Ti \over \mH} \right] + 
{d\Fcond \over dz} = \Q+e - \Lambda,
\label{eq:general}
\end{equation}
where the conduction flux due to electrons is given by
\begin{equation}
\Fcond=\kappao \Te^{5/2} {d\Te \over dz}
\end{equation}
Here we included the effect of radial advection into the cooling term and
we neglected the conduction due to ions.

Following McKee \& Begelman (1990) we multiply the Equation~\ref{eq:general} 
by $\Fcond$ and integrate 
in the vertical direction across the transition zone:
\begin{eqnarray}
\lefteqn{\int_{z_1}^{z_2}{d \over dz} \left[ \rho v_{\rm z} {5 \over 2} 
{k \Ti \over \mH} \right] \Fcond dz +
{1 \over 2} \left[ \Fcond ^2(z_2) - \Fcond ^2(z_1) \right] = }\nonumber \\
& & =\int_{\Ts}^{\Te(P)} [\Q+e - \Lambda]\kappao \Te^{5/2} d\Te 
\label{eq:eninteg}    
\end{eqnarray}

The conduction flux at the upper and lower point of the transition zone
vanishes which is used as a boundary condition in full numerical computations
(e.g. Dullemond 1999) so the second term on the left hand side of the equation
vanishes thus leaving a relation between the mass transfer (first term) and
the integral on the right hand side.

We assume that the transition from hot coronal to cold disk material in the 
vertical direction takes place in a very narrow geometrical zone so any change
of the pressure across this zone can be neglected. This allows to calculate
the right hand side integral conveniently, using algebraic expressions for
the density, ion temperature and the heating and cooling as functions of the
electron temperature.

The left hand side can be approximately expressed by estimating first the
Field length (Field 1965) measuring the thickness of the transition zone
\begin{eqnarray}
\lambdaF &=& \left({\kappao \Te^{7/2}(P) \over <|\Q+e - \Lambda>}
\right)^{1/2} \approx \nonumber \\
& \approx & {\kappao \Te^{7/2}(P) \over |\int_{\Ts}^{\Te(P)} 
[\Q+e - \Lambda]\kappao \Te^{5/2} d\Te|^{1/2}}.
\label{eq:lamfield}
\end{eqnarray}
We now  estimate the typical value of the conductive flux within the 
transition zone as
\begin{equation}
<\Fcond> = \kappao \Te (P)^{7/2}/\lambdaF
\end{equation}
which allows us to express the left hand side term as
\begin{equation}
\int_{z_1}^{z_2}{d \over dz} \left[ \rho v_{\rm z} {5 \over 2} 
{k \Ti \over \mH} \right] 
\Fcond dz \approx 
<\Fcond> \mdotz {5 \over 2} 
{k \Ti (P)\over \mH}.
\end{equation}

Therefore the approximate formula for the condensation/evaporation rate at
the disk/corona boundary reads
\begin{equation}
\mdotz = { \int_{\Ts}^{\Te(P)} [\Q+e - \Lambda]\kappao \Te^{5/2} d\Te     
\over \kappao \Te (P)^{7/2}} \lambdaF 
 \left( \frac{2}{5} \frac{\mH}{k \Ti (P)} \right) 
\label{eq:conpar}
\end{equation}

\end{document}